\documentclass[10pt,twocolumn,letterpaper]{article}

\usepackage{cvpr}              

\usepackage{url}
\usepackage{cite}
\usepackage{float}
\usepackage{graphicx}
\usepackage{amsmath}
\usepackage{amssymb}
\usepackage{array}
\usepackage{booktabs}
\usepackage{multirow}
\usepackage{diagbox}
\usepackage[T1]{fontenc}
\usepackage[table,xcdraw]{xcolor}
\usepackage[accsupp]{axessibility}

\usepackage[pagebackref,breaklinks,colorlinks]{hyperref}

\usepackage[capitalize]{cleveref}
\crefname{section}{Sec.}{Secs.}
\Crefname{section}{Section}{Sections}
\Crefname{table}{Table}{Tables}
\crefname{table}{Tab.}{Tabs.}

\def\cvprPaperID{12830} 
\def\confName{CVPR}
\def\confYear{2023}

\begin{document}

\title{Learning a Practical SDR-to-HDRTV Up-conversion using New Dataset and Degradation Models}

\author{Cheng Guo$^{1,2}$, Leidong Fan$^{3,2}$, Ziyu Xue$^{4,1}$ and Xiuhua Jiang$^{2,1}$\\
$^{1}$State Key Laboratory of Media Convergence and Communication, Communication University of China\\
$^{2}$Peng Cheng Laboratory $^{3}$Peking University\\
$^{4}$Academy of Broadcasting Science, National Radio and Television Administration\\
{\tt\small \{guocheng,jiangxiuhua\}@cuc.edu.cn, fanleidong@stu.pku.edu.cn, xueziyu@abs.ac.cn}
}

\twocolumn[{
	\maketitle
	\begin{figure}[H]
		\hsize=\textwidth
		\centering
		\includegraphics[width=2\linewidth]{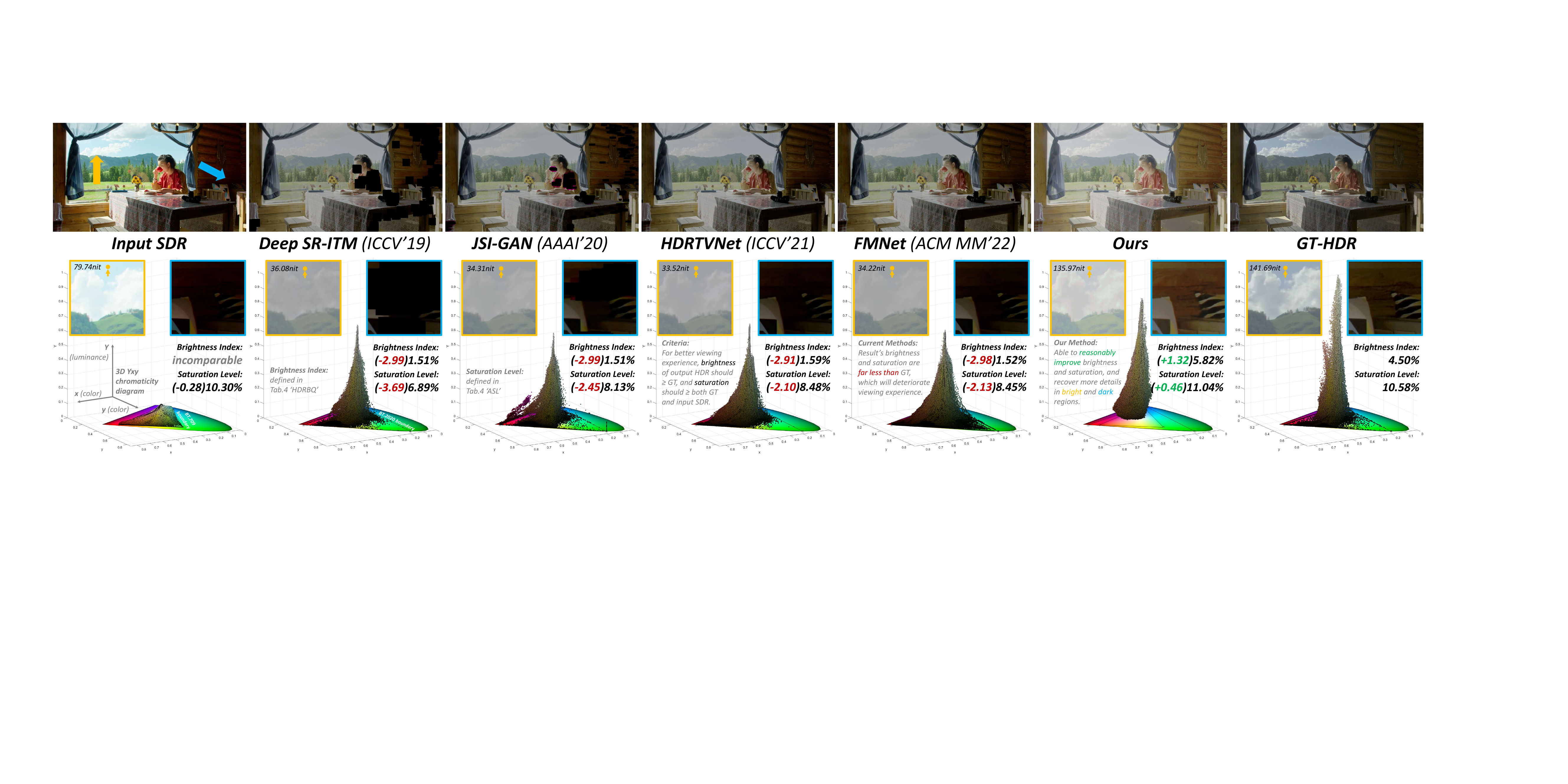}
		\caption{\textbf{Up-converting SDR content for HDR-WCG TV display}. To ensure better viewing experience\cite{BT2381}, we use (1) \textit{3D Yxy chromaticity diagram} (vertical axis \textit{Y} for HDR/lumiannce, \textit{xy} plane for WCG/color) to see how HDRTV's advance on HDR\&WCG volume is recovered, and (2) \textit{detailed visuals} (yellow and blue boxes) to assess method's recover ability. Note that HDR is dimer here in print version since its large luminance\&color container is interpreted by small SDR capacity, and will appear normal if correctly visualized (top Fig.\ref{fig:subj}). Still, from (3) `\textit{Brightness Index}' \& '\textit{Saturation Level}' we know that result from \textbf{current methods} is more \textbf{dim and desaturated} than GT.}
		\label{fig:teaser}
	\end{figure}
}]

\begin{abstract}
In media industry, the demand of SDR-to-HDRTV up-conversion arises when users possess HDR-WCG (high dynamic range-wide color gamut) TVs while most off-the-shelf footage is still in SDR (standard dynamic range).
The research community has started tackling this low-level vision task by learning-based approaches. When applied to real SDR, yet, current methods tend to produce dim and desaturated result, making nearly no improvement on viewing experience.
Different from other network-oriented methods, we attribute such deficiency to training set (HDR-SDR pair).
Consequently, we propose new HDRTV dataset (dubbed HDRTV4K) and new HDR-to-SDR degradation models.
Then, it's used to train a luminance-segmented network (LSN) consisting of a global mapping trunk, and two Transformer branches on bright and dark luminance range.
We also update assessment criteria by tailored metrics and subjective experiment. Finally, ablation studies are conducted to prove the effectiveness.
Our work is available at: \url{https://github.com/AndreGuo/HDRTVDM}.

\end{abstract}

\section{Introduction}
\label{sec:intro}

The dynamic range of image is defined as the maximum recorded luminance to the minimum.
Larger luminance container endows high dynamic range (HDR) a better expressiveness of scene.
In media and film industry, the superiority of HDR is further boosted by advanced electro-optical transfer function (EOTF) \eg PQ/HLG\cite{BT2100}, and wide color-gamut (WCG) RGB primaries \eg BT.2020\cite{BT2020}.

While WCG-HDR displays are becoming more readily available in consumer market, most commercial footage is still in standard dynamic range (SDR) since WCG-HDR version is yet scarce due to exorbitant production workflow.
Hence, there raise the demand of converting vast existing SDR content for HDRTV service. Such SDR may carry irreproducible scenes, but more likely, imperfections brought by old imaging system and transmission.
This indicates that SDR-to-HDRTV up-conversion is an ill-posed low-level vision task, and research community has therefore begun involving learning-based methods (\cite{Kim19,Kim202,Zeng20,Chen211,Cao22,Xu222} \etc).

Yet, versatile networks they use (\S \ref{sec:our_scope}), we find current methods' result dim and desaturated when feeding real SDR images (Fig.\ref{fig:teaser}), conflicting with the perceptual motive of SDR-to-HDRTV up-conversion.
As reported by CVPR22-1st Workshop on Vision Dataset Understanding\cite{VDU}, most methods are network-oriented and understate the impact of training set.
For restoration-like low-level vision, there are 2 ingredients of a training set: the quality of label GT itself, and the GT-to-LQ degradation model (DM) \ie what the network learns to restore.
Such neglect is getting remedied in other low-level vision tasks\cite{Zhang21SRDM,Dewil22DNDM,Jiang21JPEGDM,Jiang22IEDM,Zhou22UDCDM,Guo2022SIHDR}, but still pervasive in learning-based SDR-to-HDRTV up-conversion.

Not serendipitously, we find dataset the reason why current methods underperform.
We exploit several HDRTV-tailored metrics (Tab.\ref{tab:hdr-wcg-metrics}) to assess current training set:
(1) by measuring label HDR's \textit{extent of HDR/WCG} \etc (Tab.\ref{tab:hdr_stat}), we notice that its quality and diversity are inadequate to incentive the network to produce appealing result,
(2) via the \textit{statistics of degraded SDR}, we find current HDR-to-SDR DMs' tendency to exaggeratedly alter the saturation and brightness (see Tab.\ref{tab:sdr_stat}) thus network will learn a SDR-to-HDR deterioration.
Hence, we propose HDRTV4K dataset (\S \ref{sec:training_dataset}) consisting of high-quality and diversified (Fig.\ref{fig:tsne}) HDRTV frames as label.
Then exploit 3 new HDRTV-to-SDR DMs (\S \ref{sec:training_dm}) avoiding above insufficiency, meanwhile possessing appropriate degradation capability (Tab.\ref{tab:sdr_stat}) so the network can learn reasonable restoration ability.

Afterwards, we formulate the task as the combination of global mapping on the \textit{full luminance range} and recovery of \textit{low/high luminance range}.
Correspondingly, we propose Luminance Segmented Network (LSN, \S \ref{sec:net_structure}) where a global trunk and two Transformer-style UNet\cite{Restormer} branches are assigned to respectively execute divergent operations required in different segmented luminance ranges (areas).

Lastly, as found by \cite{CheatHDR,SIHDRQA}, conventional distance-based metrics well-performed in solely-reconstruction task (\eg denoising) fail for perceptual-motivated HDR reconstruction, we therefore update the assessment criteria with fine-grained metrics (\S \ref{sec:exp_metrics}) and subjective experiment (\S \ref{sec:exp_subj}) \etc.

Our contributions are three-fold:
\textbf{(1)} Emphasizing \& verifying the impact of dataset on SDR-to-HDRTV task, which has long been understated.
\textbf{(2)} Exploiting novel HDRTV dataset and HDR-to-SDR degradation models for network to learn.
\textbf{(3)} Introducing new problem formulation, and accordingly proposing novel luminance segmented network.

\section{Related Works}
\label{sec:related_works}

\subsection{Nomenclature and our scope}
\label{sec:our_scope}

Plenty of HDR-related learning-based methods \cite{HDRDNNSurvey} have been proposed, yet, they markedly differ in intended application.
As in Tab.\ref{tab:nomenclature}, `linear HDR' means \textit{scene-referred} HDR images dedicating to record the linear radiance for graphics application \eg image-based lighting\cite{IBL,ReinhardHDRBook}, and `HDRTV' stands for our \textit{display-referred} WCG-HDR\cite{BT2100} format.
Clearly, \textcircled{1} synthesize HDR view on SDR display, \textcircled{2} emphasize dealing with misalignment between multiple SDRs and is designed for new imaging pipeline, while \textcircled{3} and \textcircled{4} are all oriented for existing SDR, but for different application.
Since many works are proposed before community's clarity on above nomenclature, we classify a method as SDR-to-HDRTV up-conversion (\textcircled{4}, our task) if its target HDR is claimed in PQ/HLG EOTF and BT.2020 WCG.

\begin{table}[!h]
	\footnotesize
	\centering
	\begin{tabular}{|c|c|c|c|c|}
		\hline
		& \textbf{Task name}                                                   & \textbf{From}                                                         & \textbf{To} single                                                & \textbf{Methods} \\ \hline
		\textcircled{1}                                                  & \begin{tabular}[c]{@{}c@{}}HDR-style\\ enhancement\end{tabular}      & \begin{tabular}[c]{@{}c@{}}single\\ SDR\end{tabular}                  & \begin{tabular}[c]{@{}c@{}}enhanced\\ SDR\end{tabular}                & \eg \cite{Gharbi17HDRIE,Yang18HDRIE,Zheng21HDRIE,Mildenhall22HDRIE}    \\ \hline
		\textcircled{2}                                                  & \begin{tabular}[c]{@{}c@{}}multi-exposure\\ HDR imaging\end{tabular} & \begin{tabular}[c]{@{}c@{}}multiple\\ SDR\end{tabular}                & \multirow{2}{*}{\begin{tabular}[c]{@{}c@{}}\textcolor[HTML]{009999}{linear} \\ \textcolor[HTML]{009999}{HDR}\end{tabular}} & \eg \cite{Kalantari18MEHDR,Wu18MEHDR,Yan19MEHDR,Chen21MEHDR,Niu21MEHDR,NTIRE22MEHDR}    \\ \cline{1-3} \cline{5-5} 
		\textcircled{3}                                                  & \textcolor[HTML]{669933}{SI-HDR}$^\star$                                 & \multirow{3}{*}{\begin{tabular}[c]{@{}c@{}}single\\ SDR\end{tabular}} &                                                                       & \eg \cite{Eilertsen17SIHDR,Marnerides18SIHDR,Liu20SIHDR,Santos20SIHDR,Chen21SIHDR}   \\ \cline{1-2} \cline{4-5} 
		\begin{tabular}[c]{@{}c@{}}\textcircled{4}\\ (ours)\end{tabular} & \begin{tabular}[c]{@{}c@{}}\textcolor[HTML]{669933}{iTM}$^\star$ or\\ \textcolor[HTML]{009999}{up-convertion}\end{tabular}       &                                                                       & \begin{tabular}[c]{@{}c@{}}\textcolor[HTML]{666699}{HDRTV}\\ frame\end{tabular}                 & \begin{tabular}[c]{@{}c@{}}\cite{Kim19,Kim202,Zeng20,Chen211,Cao22,Xu222}\\ \cite{Ning18,Hirao18,Kim18,Xu191,Xu192,Kim201,Zou20,Chen212,Xu221,He221,He222,Xu223,Shao22,Mustafa22,Yao23,Tang22}\end{tabular}    \\ \hline
		\multicolumn{5}{l}{\scriptsize $^\star$: SI-HDR: Single-Image HDR reconstruction, iTM: inverse Tone-Mapping.} \\
	\end{tabular}
	\caption{Various learning-based HDR-related tasks. \textcolor[HTML]{669933}{Green}/\textcolor[HTML]{009999}{cyan}/ \textcolor[HTML]{666699}{magenta} terms are respectively from \cite{CheatHDR}/our/\cite{Chen211} nomenclature.}
	\label{tab:nomenclature}
\end{table}
 
 In our scope (\textcircled{4}), methods designed their networks with distinct philosophy:
 Except \eg semantic-requiring highlight recovery, main process of our task belongs to global mapping resembling image retouching/enhancement.
 Following this, \cite{Chen211} formulates the task as 3 steps and use $1\times1$ convolution for global part, \cite{Kim201} learn the mapping between small sorted image patches, while \cite{Mustafa22} conducts explicitly-defined per-pixel mapping using the learned latent vector.
 
 Feature modulation is also popular to involve global prior information. Compared with \cite{Chen211}, they change:
 prior's type\cite{He221,Xu223}, modulation tensor's shape\cite{He222,Shao22,Xu223}, modulation mechanism\cite{Xu222} and modulation's position \cite{Shao22}.
 
 To coordinate consecutive video frames, \cite{Xu192} applies 3D-convolution with extra temporal-D, \cite{Cao22} take 3 frames as input and deploy multi-frame interaction module at UNet bottleneck. Also, for alignment, \cite{Zou20,Xu223} use deformable convolution whose offset map is predicted by different modules.
 
 Due to the resolution discrepancy between SDRTV and HDRTV, some methods\cite{Kim19,Kim202,Zeng20,Xu221,He222,Yao23} jointly conduct super-resolution and up-conversion. 
 Also, some methods are assisted by non-learning pre-processing\cite{Tang22} or bypass\cite{Kim201,Chen212}, while \cite{Xu191} is trained for badly-exposed SDR. 

\subsection{HDRTV dataset}
\label{sec:hdrtv_dataset}

Diversified networks they use, there're currently only 3 open HDRTV training set (Tab.\ref{tab:hdrtv_dataset}). All datasets (including ours) contain HDRTV frames in (D65 white) BT.2020\cite{BT2020} RGB primaries (gamut), PQ\cite{ST2084} EOTF and 1000$nit$ peak luminance, and SDR counterpart in BT.709\cite{BT709} gamut, \textit{gamma} EOTF and 100$nit$ peak luminance.

\begin{table}[!h]
	\centering
	\scriptsize
	\begin{tabular}{|l|r|c|c|}
		\hline
		\textbf{Dataset} (\textbf{Usage}) & \textbf{\#pair} & \textbf{Resolution} & \textbf{HDR format}    \\ \hline
		KAIST\cite{Kim19} (\cite{Kim202,He222,Yao23})   & 39840   & 160$\times$160    & {\tiny uint16 MATLAB .mat YUV}  \\ \hline
		Zeng20\cite{Zeng20}   & 23229   & \multirow{2}{*}{UHD}  & {\tiny H.265 main10 YUV} \\ \cline{1-2} \cline{4-4} 
		HDRTV1K\cite{Chen211} (\cite{He221,Xu222,Shao22})    & 1235    &            & {\tiny 16bit .png RGB}   \\ \hline
		HDRTV4K (ours new)     & 3878    & HD\&UHD     & {\tiny 16bit lossless .tif RGB}   \\ \hline
	\end{tabular}
	\caption{Status of different HDRTV dataset: training set part. Our HDR frames sized both HD (1920$\times$1080) and UHD (3840$\times$2160) are manually chosen from >220 different videos clips, and encapsulated in lossless (\textit{LZW} or \textit{deflate}) TIFF. Others are respectively extracted from only 7\cite{Kim19}/18\cite{Chen211} TV demos and 1 graded movie\cite{Zeng20}, their quality and diversity are quantified later in Tab.\ref{tab:hdr_stat} \& Fig.\ref{fig:tsne}.}
	\label{tab:hdrtv_dataset}
\end{table}

\subsection{HDRTV-to-SDR degradation model}
\label{sec:hdrtv_dm}

Most methods follow the common practice to degrade label HDR(GT) to input SDR(LQ).
Degradation model (DM) matters since it determined what network can learn, and is relatively paid more attention even in SI-HDR\cite{Eilertsen17SIHDR,Liu20SIHDR,Guo2022SIHDR}.
Yet, few discussion is made in SDR-to-HDRTV (Tab.\ref{tab:hdrtv_dm}):

\begin{table}[!h]
	\centering
	\scriptsize
	\begin{tabular}{|c|c|c|c|}
		\hline
		\textbf{DM} & \textit{\textbf{YouTube}} & \textit{\textbf{Reinhard}} & other DMs \\ \hline
		Usage & \cite{Kim18,Kim19,Kim202,Zou20,Chen211,Cao22,Xu221,He221,He222,Xu222,Xu223,Shao22,Yao23} & \cite{Ning18,Xu191,Xu192,Kim201,Zeng20} & \multirow{2}{*}{\begin{tabular}[c]{@{}c@{}} \textbf{\textit{2446a}}\cite{Chen212}\\ \etc\cite{Mustafa22,Tang22,Cheng22ITMDM} \end{tabular}} \\ \cline{1-3}
		Dataset & KAIST \& HDRTV1K & Zeng20            &                  \\ \hline
	\end{tabular}
	\caption{Current HDRTV-to-SDR degradation models (DMs). `Dataset' means SDR there is degraded from HDR using that DM.}
	\label{tab:hdrtv_dm}
\end{table}

\textit{\textbf{Youtube}} stands for the default conversion YouTube applied to BT.2020/PQ1000 HDR content to produce its SDR-applicable version, \textit{\textbf{Reinhard}}/\textit{\textbf{2446a}} means tone-mapping HDR to SDR using \textit{Reinhard TMO}\cite{ReinhardTMO}/BT.2446\cite{BT2446}\textit{Method A}. \cite{Mustafa22}/\cite{Tang22}/\cite{Cheng22ITMDM} respectively degrade HDR to SDR by grading/\textit{Habel TMO}/another learned network.

In other restoration tasks\cite{Zhang21SRDM,Dewil22DNDM,Jiang21JPEGDM,Jiang22IEDM,Zhou22UDCDM,Guo2022SIHDR}, DMs are designed to have proper extent and diversity of degradation so network can learn appropriate restore capability and good generalization.
Accordingly, we argue that current DMs are not favorable for training.
Specifically, the motive of \textit{\textbf{YouTube}} is to synthesize HDR view for user possessing only SDR display, it tends to enhance/increase the brightness and saturation so network will, vise-versa, learn to deteriorate/decline them.
Also, tone-mapping \eg \textit{\textbf{Reinhard}} and \textit{\textbf{2446a}} dedicate to preserve as much information from HDR, they have monotonically increasing mapping curves (Fig.\ref{fig:dm_curve}) without highlight clipping, therefore the trained network is not likely to recover much information in over-exposed areas. Above observations are later proven in Tab.\ref{tab:sdr_stat}, Fig.\ref{fig:teaser}\&\ref{fig:result} \etc.

\section{Proposed Method}
\label{sec:method}

\begin{figure*}[!h]
	\centering
	\includegraphics[width=0.9\linewidth]{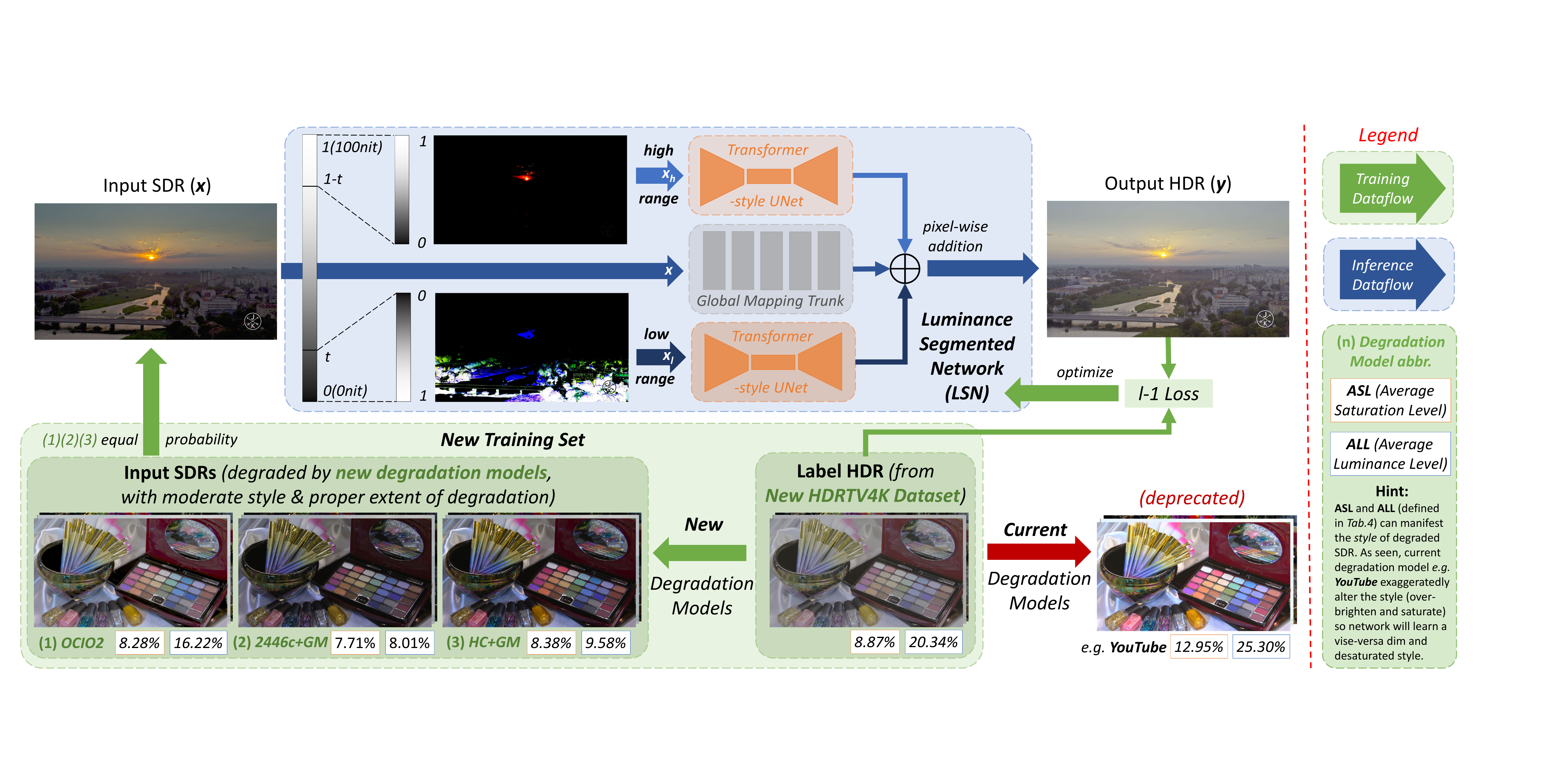}
	\caption{Overview of this work. Our Luminance Segmented Network (LSN, \S \ref{sec:network}) is designed based on novel problem fromulation, then supervisedly trained with label HDR from the proposed HDRTV4K dataset (\S \ref{sec:training_dataset}), and input SDR degraded by novel degradation models (DMs, \S \ref{sec:training_dm}). The \textbf{major concerns} of our LSN, HDRTV4K dataset, and DMs are respectively: recovering dark\&bright areas, improving the quality (Tab.\ref{tab:hdr_stat}) and diversity (Fig.\ref{fig:tsne}) of label GT-HDR, and ensuring the LQ-SDR is with proper style and degradation (Tab.\ref{tab:sdr_stat} \& Fig.\ref{fig:dm_curve}).}
	\label{fig:overview}
\end{figure*}

The overview of our method is illustrated in Fig.\ref{fig:overview}.

\subsection{Network structure}
\label{sec:network}

\begin{figure}[!h]
	\centering
	\includegraphics[width=\linewidth]{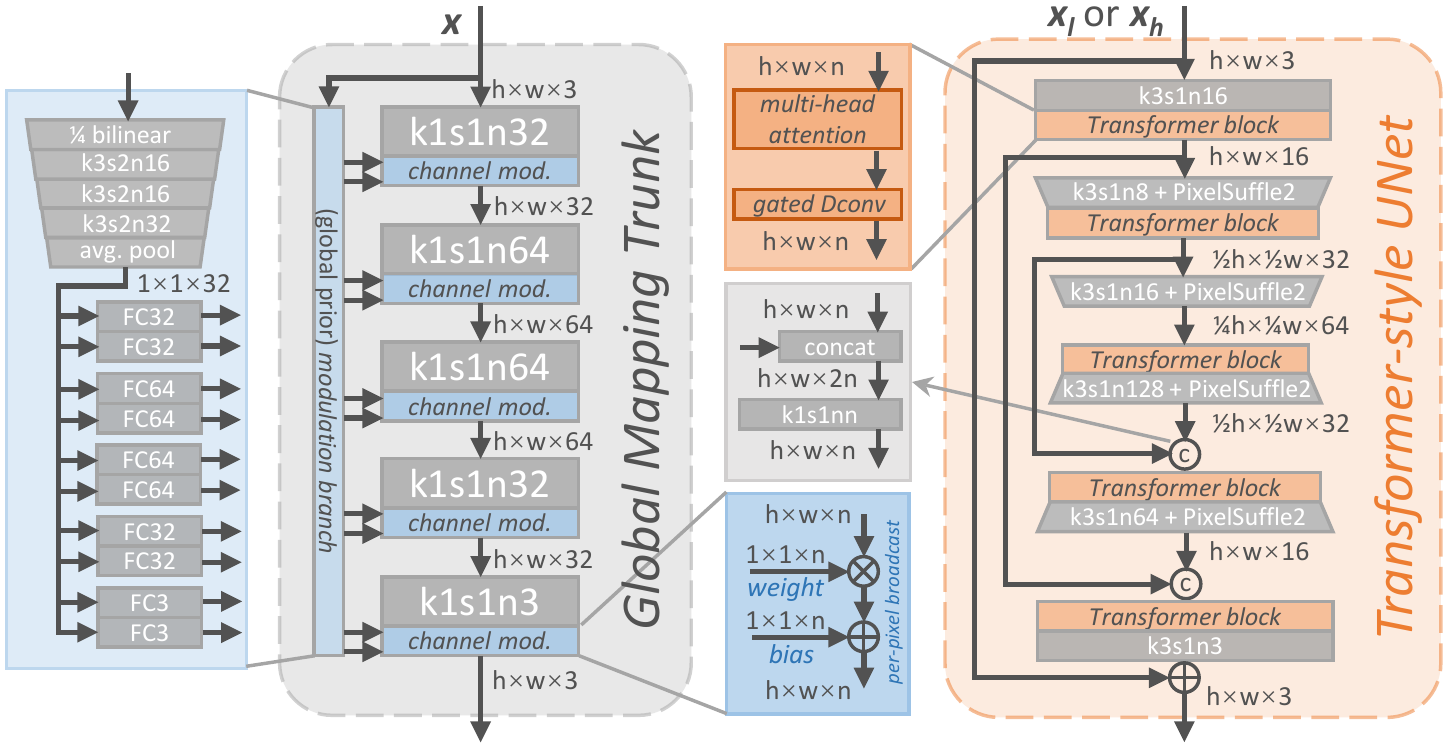}
	\caption{Detailed structure of LSN modules. `k3s1n16' means convolution layer with 3$\times$3 {\tt kernel}, {\tt stride}=1, {\tt out nc}=16, and `FC32' is fully-connected layer with {\tt out nc}=32. We deploy only 1 Transformer block\cite{Restormer} at each en/decode level of UNet to lighten LSN (\#param=325k) for its application on 4K resolution.}
	\label{fig:network}
\end{figure}

\label{sec:formulation}
\textbf{Problem formulation} helps the researcher to clarify what degradation the network should restore. Given $\mathbf{x}$ the SDR, $\mathbf{y}$ the HDR.
Previous methods\cite{Chen211,He221,Xu223} hypothesis that $\mathbf{x}$ and $\mathbf{y}$ are from single RAW file of same camera.

This applies to SDR-HDR synchronous production where $\mathbf{x}$ is also of high quality, thus up-conversion ($\mathbf{y}=f(\mathbf{x})$) is only supposed to follow specific style and recover less missing information.
However, since our method is for existing SDR, we assume that SDR ($\mathbf{x}$) and it imaginary HDR counterpart ($\mathbf{y}$) were simultaneously shot by different camera and later pipeline, similar to \cite{CheatHDR}.
The imperfection of SDR pipeline makes existing SDR susceptible to limited latitude, narrow gamut and other degradations ($\mathbf{x}=d(\mathbf{y})$).

Some works\cite{Chen211,Liu20SIHDR} suppose $d(\cdot)$ are introduced orderly, and assign hierarchical sub-networks to for $f(\cdot)$.
Since cascaded networks are bulky for real application, we instead formulate that specific degradation is more visible in different \textit{luminance range}.
Concretely, over-exposure occurs in \textit{high-luminance range} in SDR ($\mathbf{x}_h$), \textit{low-luminance range} ($\mathbf{x}_l$) is more susceptible to noise \etc, while the quality of \textit{mid-luminance range} ($\mathbf{x}_m$) is relatively higher.
Segmented luminance ranges are treated separately by some traditional up-conversion operators (\cite{Mohammadi20} \etc), and here we introduce this idea to our deep neural network (DNN)---LSN:

\label{sec:net_structure}
\textbf{Luminance Segmented Network (LSN)} consist of an trunk on \textit{full-luminance range} ($\mathbf{x}$), and 2 branches respectively on $\mathbf{x}_l$ and $\mathbf{x}_h$ which are segmented by:
\begin{equation}
	\mathbf{x}_l = max(0,\frac{t-\mathbf{x}}{t}),\ \mathbf{x}_h = max(0,\frac{\mathbf{x}-1}{t}+1)
	\label{eq:lum_seg_mask}
\end{equation}
where $\mathbf{x}\in[0,1]$. That is, luminance range \textit{lower}/\textit{higher} than threshold $t$ (empirically set to $0.05$) is linearly mapped to more significant value $[0,1]$, as in top-right Fig.\ref{fig:overview}.

After segmentation, $\mathbf{x}$, $\mathbf{x}_l$ and $\mathbf{x}_h$ require distinct DNN operation.
First, as found by \cite{Chen211} \etc (\S \ref{sec:our_scope}), the majority of $f(\cdot)$ belongs to global (pixel-independent) operation similar to image enhancement/retouching.
Therefore, we assign 5 cascaded receptive-field-free $1\times1$ convolution layers on \textit{full luminance range} ($\mathbf{x}$) as the \textbf{Global Mapping Trunk}.
Similar to \cite{Chen211,Xu222,He221,He222,Xu223,Shao22}, we append a modulation branch (Fig.\ref{fig:network} left) to involve $\mathbf{x}$'s global prior into DNN's decision.

On the other hand, $f(\cdot)$ still involves non-global restoration where local semantic should be aggregated.
Specifically, DNN's mission in $\mathbf{x}_l$ is similar to denoising and low-light enhancement, while that in $\mathbf{x}_h$ resembles image inpainting (claimed by \cite{Marnerides21SIHDR,Zhang21SIHDR}) where lost content is hallucinated.
These low-level vision tasks require larger receptive-filed, we hence choose Transformer blocks\cite{Restormer} specializing long-distance dependency, and arranged them as encoder-decoder with skip-connections (UNet).
\textbf{Transformer-style UNet branches} on $\mathbf{x}_l$ and $\mathbf{x}_h$ share the same structure but different parameters.
The detail of LSN is depicted in Fig.\ref{fig:network}.

\subsection{HDRTV4K dataset}
\label{sec:training_dataset}

After designing LSN, we need better training set to unleash its capability. We start with the quality of label HDR:

\textbf{Motivation.} In \S\ref{sec:intro}, we argue the quality of label HDR in current datasets. Here, it is assessed from 3 aspects by 10 metrics in Tab.\ref{tab:hdr-wcg-metrics}, based on following considerations:

\begin{table}[!h]
	\centering
	\scriptsize
	\begin{tabular}{|cc|}
		\hline
		\multicolumn{2}{|c|}{\cellcolor[HTML]{CCCCCC}{Metrics on the \textit{extent of HDR/WCG} $\downarrow$}} \\ \hline
		\multicolumn{1}{|c|}{\textbf{FHLP}}                 & \begin{tabular}[c]{@{}c@{}}\textbf{F}raction of \textbf{H}igh\textbf{L}ight \textbf{P}ixel: Spatial ratio of `highlight' pixel \\ \ie whose normalized luminance $Y=0.2627R+0.6780G$ \\ $+0.0593B>0.1$ (100$nit$, SDR's peak luminance.)\end{tabular}          \\ \hline
		\multicolumn{1}{|c|}{\textbf{EHL}} & \begin{tabular}[c]{@{}c@{}}\textbf{E}xtent of \textbf{H}igh\textbf{L}ight: Average pxiel ($i$) distance between\\ the luminance of HDR and its clip-to-100$nit$ version$^1$: \\ $\frac{1}{n} {\textstyle {\sum_{i=1}^{n}} \sqrt{[Y_i-clip(Y_i)]^2}},\ clip(x) = clamp(x,0,0.1)$\end{tabular} \\ \hline
		\multicolumn{1}{|c|}{\textbf{FWGP}}                 & \begin{tabular}[c]{@{}c@{}} \textbf{F}raction of \textbf{W}ide-\textbf{G}amut \textbf{P}ixel: Spatial ratio of WCG pixel \ie \\ whose $[x,y]$ coordinates fall inside BT.2020 but outside\\ SDR's BT.709 gamut in $Yxy$ chromaticity diagram. \end{tabular}          \\ \hline
		\multicolumn{1}{|c|}{\textbf{EWG}} & \begin{tabular}[c]{@{}c@{}} \textbf{E}xtent of \textbf{W}ide-\textbf{G}amut\cite{EWG}: Average pixel-distance between \\ WCG-HDR and its gamut-hard-clipped\cite{BT2407} version$^2$: \\ $\frac{1}{n} {\textstyle {\sum_{i=1}^{n}} {\left \| \mathbf{S}_i-HC(\mathbf{S}_i) \right \|}_2},\ \mathbf{S} = [X,Y,Z]^\mathrm{T}$ \end{tabular}          \\ \hline
		\multicolumn{2}{|c|}{\cellcolor[HTML]{CCCCCC}{Metrics on \textit{intra-frame diversity} $\downarrow$ (all variance-based)}} \\ \hline
		\multicolumn{1}{|c|}{\textbf{SI}} & \begin{tabular}[c]{@{}c@{}} \textbf{S}patial \textbf{I}nformation:
			Standard deviation over the pixels of\\ Sobel-filtered frame, defined in Annex 6 of \cite{BT500}. \end{tabular}          \\ \hline
		\multicolumn{1}{|c|}{\textbf{CF}} & \textbf{C}olor\textbf{F}ulness: Defined in \cite{CF}.  \\ \hline
		\multicolumn{1}{|c|}{\textbf{stdL}} & standard deviation of \textbf{L}uminance, over all pixels of a frame. \\ \hline
		\multicolumn{2}{|c|}{\cellcolor[HTML]{CCCCCC}{Metrics on \textit{overall-style} $\downarrow$}} \\ \hline
		\multicolumn{1}{|c|}{\textbf{ASL}} & \begin{tabular}[c]{@{}c@{}} \textbf{A}verage \textbf{S}aturation \textbf{L}evel: Normalized pixel-average length \\ of HDR chrominance component $\mathbf{C}=[C_t,C_p]^\mathrm{T}$\cite{ICtCp}: \\ $\frac{\sqrt{2}}{n} {\textstyle {\sum_{i=1}^{n}} \left \| \mathbf{C}_i \right \|_2}, \ \mathbf{C} \in [-0.5,0.5]$ \end{tabular}          \\ \hline
		\multicolumn{1}{|c|}{\textbf{ALL}} & \textbf{A}verage \textbf{L}uminance \textbf{L}evel: Pixel-average of $Y$ in \textbf{FHLP}.  \\ \hline
		\multicolumn{1}{|c|}{\textbf{HDRBQ}} & \textbf{HDR} \textbf{B}rightness \textbf{Q}uantification\cite{HDRBQ}, visual salience involved. \\ \hline
		\multicolumn{2}{l}{\tiny $^1$: \textbf{EHL} is to compensate cases \eg an all-101$nit$ HDR frame with 100\% \textit{FHLP} but less extent of highlight.} \\
		\multicolumn{2}{l}{\tiny $^2$: You can find the formulation of gamut hard-clipping ($HC(\cdot)$) at Eq.\ref{eq:gm_hc}.} \\
	\end{tabular}
	\caption{The quality and diversity of label HDR is measured from 3 aspects by 10 metrics above (both in positive correlation), results are in Tab.\ref{tab:hdr_stat}. See supplementary material for full illustration.}
	\label{tab:hdr-wcg-metrics}
\end{table}


First, greater \textit{extent of HDR/WCG} stands more probability for network to learn pixel in advanced WCG-HDR volume beyond SDR's capability.
Meanwhile, higher \textit{intra-frame diversity} means better generalization the network can learn within small batch-size, \ie bigger \textbf{SI}/\textbf{CF}/\textbf{stdL} indicate more diversified high-frequency patterns/richer color volume/greater luminance contrast for network to learn.

Also, style distillation\cite{Mustafa22} has become a major contributor to method's performance in similar task \eg image enhancement/retouching, we therefore append 3 metrics quantifying the \textit{overall-style} of label HDR. Note that network's learned style will be, substantially, affected more by degradation model (DM) which will be discussed later.

\begin{table*}[!h]
	\centering
	\small
	\begin{tabular}{|l|rr|rr|rrr|rrr|}
		\hline
		& \multicolumn{2}{c|}{\textit{Extent of HDR}}                                 & \multicolumn{2}{c|}{\textit{Extent of WCG}}                                & \multicolumn{3}{c|}{\textit{Intra-frame diversity}}     & \multicolumn{3}{c|}{\textit{Overall-style}}                                                                                                                                           \\ \cline{2-11} 
		\multirow{-2}{*}{\diagbox{\textbf{Dataset}}{\textbf{Metrics}}} & \multicolumn{1}{c}{\textbf{FHLP}}       & \multicolumn{1}{c|}{\textbf{EHL}} & \multicolumn{1}{c}{\textbf{FWGP}}       & \multicolumn{1}{c|}{\textbf{EWG}} & \multicolumn{1}{c}{\textbf{SI}} & \multicolumn{1}{c}{\textbf{CF}}   & \multicolumn{1}{c|}{\textbf{stdL}} & \multicolumn{1}{c}{\textbf{ASL}} & \multicolumn{1}{c}{\textbf{ALL}} & \multicolumn{1}{c|}{\textbf{HDRBQ}}     \\ \hline
		KAIST\cite{Kim19}              & 1.5250                         & 0.2025                         & \cellcolor[HTML]{EEFFEE}5.4771 & 0.1104                         & 1.9372                         & 5.9485                          & 0.9597                        & 8.9087                  & 17.2854                 & 1.8597                          \\
		Zeng20\cite{Zeng20}            & 0.0197                         & 0.0012                         & 0.4792                         & 0.0034                         & 0.1231                         & 4.2048                          & 0.3146                        & 3.8061                  & 6.0805                  & 0.3781                          \\
		HDRTV1K\cite{Chen211}          & 1.2843                         & 0.1971                         & 2.9089                         & 0.1633                         & 2.2378                         & \cellcolor[HTML]{EEFFEE}11.0722  & 1.8006                        & 10.9414                 & 15.1626                 & 2.7970                          \\
		HDRTV4K (ours)            & \cellcolor[HTML]{EEFFEE}5.3083 & \cellcolor[HTML]{EEFFEE}0.9595 & 2.6369                         & \cellcolor[HTML]{EEFFEE}0.5123 & \cellcolor[HTML]{EEFFEE}3.5508 & 10.5882                         & \cellcolor[HTML]{EEFFEE}3.4837 & 9.8274                  & 21.1996                 & 5.1593 \\ \hline
	\end{tabular}
	\caption{Quality of label HDR frames, manifested in the frame-average of 10 metrics from Tab.\ref{tab:hdr-wcg-metrics}. Greater \textit{extent of HDR/WCG} encourages network to produce more pixels in non-SDR volume, while \textit{intra-frame diversity} is helpful for network's generalization ability. All numbers are in percentage ($\%$), and we \colorbox[HTML]{EEFFEE}{highlight} those the most favorable to training. Their diversity is further demonstrated in Fig.\ref{fig:tsne}.}
	\label{tab:hdr_stat}
\end{table*}

\textbf{Our work.} Statistics in Tab.\ref{tab:hdr_stat} confirms the deficiency of current datasets, \ie lower \textit{intra-frame diversity}, and most importantly, less \textit{extent of HDR/WCG} preventing their network from producing true HDR-WCG volume.
To this end, we propose a new dataset HDRTV4K consisting of 3878 BT.2020/PQ1000 (Tab.\ref{tab:hdrtv_dataset}) HDR frames with higher quality.

\begin{figure}[!h]
	\centering
	\includegraphics[width=\linewidth]{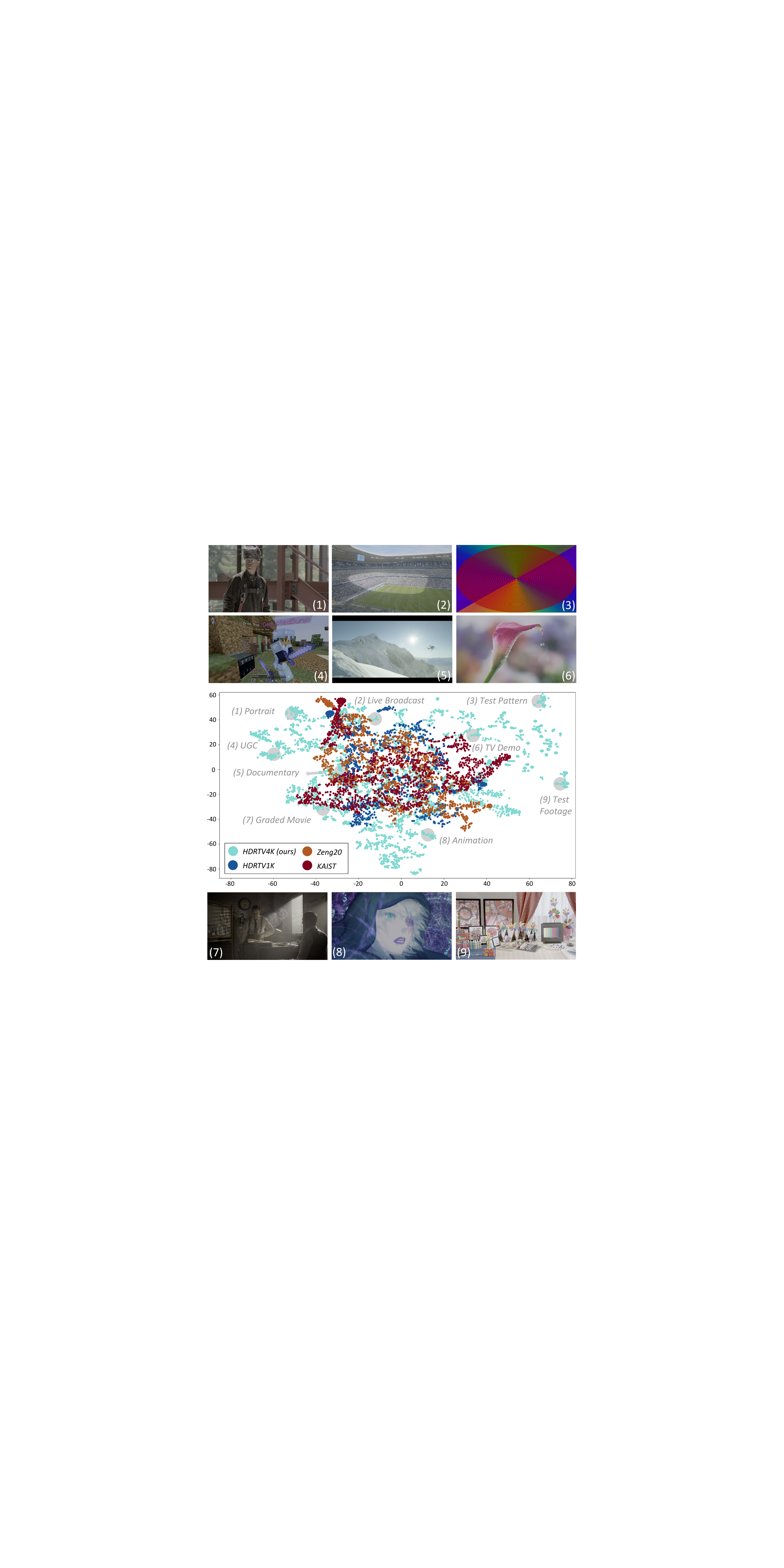}
	\caption{Diversity comparison: our HDRTV4K dataset \vs others. Here, each 2D-coordinate is the projection of single frame's 10-D vector (containing 10 metrics from Tab.\ref{tab:hdr_stat}) using t-SNE\cite{t-SNE} ({\tt exg=3}, {\tt prep=50}). We also depict thumbnail HDR frames from our dataset (1)-(9) with their corresponding 2D-coordinates highlighted with \textcolor{gray}{gray circle}. As seen, our dataset provides wider frame distribution \ie more diversified scenes for network to learn.}
	\label{fig:tsne}
\end{figure}

Specifically, these frames are manually extracted and aligned from various open content (\cite{Arri-HDR,Netflix-HDR,HdM-HDR,YouTube-UGC-HDR,LIVE-HDR-IQA} \etc) with greater \textit{extent of HDR/WCG}, higher \textit{intra-frame diversity} and reasonable \textit{style} (Tab. \ref{tab:hdr_stat}).
Note that we re-grade some original footage using DaVinci Resolve to add some perturbation on the diversity.
The thumbnails of our dataset and its frame distribution comparison are visualized in Fig.\ref{fig:tsne}.

\subsection{New degradation models}
\label{sec:training_dm}

After obtaining label HDR with higher quality, we focus on degradation model (DM) which determines the restoration network will learn.
As claimed in \S\ref{sec:hdrtv_dm}, current DMs fail for undue saturation \& brightness change and meager over-exposure degradation, which are verified by \textbf{ALL}, \textbf{ASL}, \textbf{FOEP} in Tab.\ref{tab:sdr_stat}.
That is, \textit{\textbf{YouTube}}/\textit{\textbf{2446a}} tend to increase/decline both \textbf{ASL} \& \textbf{ALL} so network will learn to simultaneously de/over-saturate and unduly-darken/brighten.
Also, \textit{\textbf{2446a}} and \textit{\textbf{Rienhard}} provide small \textbf{FOEP} which is adverse to network‘s over-exposure hallucination ability.

\begin{table}[!h]
	\centering
	\scriptsize
	\begin{tabular}{|cc|rrrrr|}
		\hline
		\multicolumn{2}{|c|}{SDR Degraded by}                                                                               & \multicolumn{1}{c}{\textbf{FOEP}$^1$}      & \multicolumn{1}{c}{\textbf{ALL}}        & \multicolumn{1}{c}{\textbf{ASL}$^2$}        & \multicolumn{1}{c}{\textbf{SI}} & \multicolumn{1}{c|}{\textbf{CF}} \\ \hline
		\multicolumn{1}{|c|}{}                                                                      & \textit{\textbf{2446a}}    & \cellcolor[HTML]{FFEEEE}0.181 & \cellcolor[HTML]{FFEEEE}5.880  & \cellcolor[HTML]{FFEEEE}3.624  & 6.041                  & 6.573                  \\
		\multicolumn{1}{|c|}{}                                                                      & \textit{\textbf{Reinhard}} & \cellcolor[HTML]{FFEEEE}1.264 & \cellcolor[HTML]{FFFFEE}21.887 & \cellcolor[HTML]{FFFFEE}7.442  & 14.147                 & 12.568                 \\
		\multicolumn{1}{|c|}{\multirow{-3}{*}{\begin{tabular}[c]{@{}c@{}}Cur-\\ rent\\DM\end{tabular}}} & \textit{\textbf{YouTube}}  & \cellcolor[HTML]{EEFFEE}5.439 & \cellcolor[HTML]{FFEEEE}28.219 & \cellcolor[HTML]{FFEEEE}14.641 & 18.545                 & 25.225                 \\ \hline
		\multicolumn{1}{|c|}{}                                                                      & \textit{\textbf{2446c+GM}} & \cellcolor[HTML]{FFFFEE}1.739 & \cellcolor[HTML]{FFFFEE}11.669 & \cellcolor[HTML]{EEFFEE}10.183  & 12.503                 & 19.391                 \\
		\multicolumn{1}{|c|}{}                                                                      & \textit{\textbf{HC+GM}}    & \cellcolor[HTML]{EEFFEE}4.252 & \cellcolor[HTML]{EEFFEE}14.062 & \cellcolor[HTML]{EEFFEE}10.377 & 15.090                 & 20.146                 \\
		\multicolumn{1}{|c|}{\multirow{-3}{*}{\begin{tabular}[c]{@{}c@{}}Ours\\ DM\end{tabular}}}   & \textit{\textbf{OCIO2}}    & \cellcolor[HTML]{FFFFEE}1.580 & \cellcolor[HTML]{EEFFEE}18.887 & \cellcolor[HTML]{EEFFEE}9.977  & 13.578                 & 18.052                 \\ \hline
		\multicolumn{2}{|c|}{criteria$\uparrow$: better when} & \multicolumn{1}{c}{kept$^3$} & \multicolumn{1}{c}{samller} & \multicolumn{1}{c}{kept} & \multicolumn{1}{c}{-} & \multicolumn{1}{c|}{-} \\ \hline
		\multicolumn{2}{|c|}{Sourse HDR (Tab.\ref{tab:hdr_stat})$^4$} & 5.308 & \textcolor[HTML]{999999}{21.200} & 9.827 & \textcolor[HTML]{999999}{3.551} & \textcolor[HTML]{999999}{10.588} \\ \hline
		\multicolumn{7}{l}{\tiny $^1$: \textbf{FOEP}: \textbf{F}raction of \textbf{O}ver-exposed \textbf{P}ixels: Spatial ratio of pixels whose normalized lumiannce $Y$= 1.} \\
		\multicolumn{7}{l}{\tiny $^2$: For SDR, \textbf{ASL} is also calculated as Tab.\ref{tab:hdr-wcg-metrics}, but with $\mathbf{C}=[C_b,C_r]^\mathrm{T}$\cite{BT709} rather $[C_t,C_p]^\mathrm{T}$.} \\
		\multicolumn{7}{l}{\tiny $^3$: Assuming that HDR's highlight (>100$nit$) part should all be clipped to 1 (over-exposure) in SDR.} \\
		\multicolumn{7}{l}{\tiny $^4$: \textcolor[HTML]{999999}{Gray} means container discrepancy \ie HDR/SDR's metric should be different even for well-degraded SDR.} \\
		\end{tabular}
	\caption{Given label HDR from HDRTV4K dataset, SDR's statistics alters when degraded by different DMs. Results are in percentage, we mark those \colorbox[HTML]{FFEEEE}{unfavorable}/\colorbox[HTML]{FFFFEE}{neutral}/\colorbox[HTML]{EEFFEE}{beneficial} for training based on the observation in \S\ref{sec:hdrtv_dm} \& \ref{sec:training_dm}. As seen, our DMs provide adequate over-exposure (\textbf{FOEP}) and no undue brightness (\textbf{ALL}) \& saturation (\textbf{ASL}) change for network to learn. Example of different degraded SDR can be found in supplementary material.}
	\label{tab:sdr_stat}
\end{table}

This motivate us to utilize/exploit new DMs with proper extent of degradation and no deficiencies on style:

(1) \textit{\textbf{OCIO2}}: \textit{OpenColorIO v2}\cite{OCIO2}, commercial method from BT.2020/PQ1000 HDR container to BT.709/\textit{gamma} SDR, implemented by 3D look-up table (LUT) here.

\begin{figure}[!h]
	\centering
	\includegraphics[width=\linewidth]{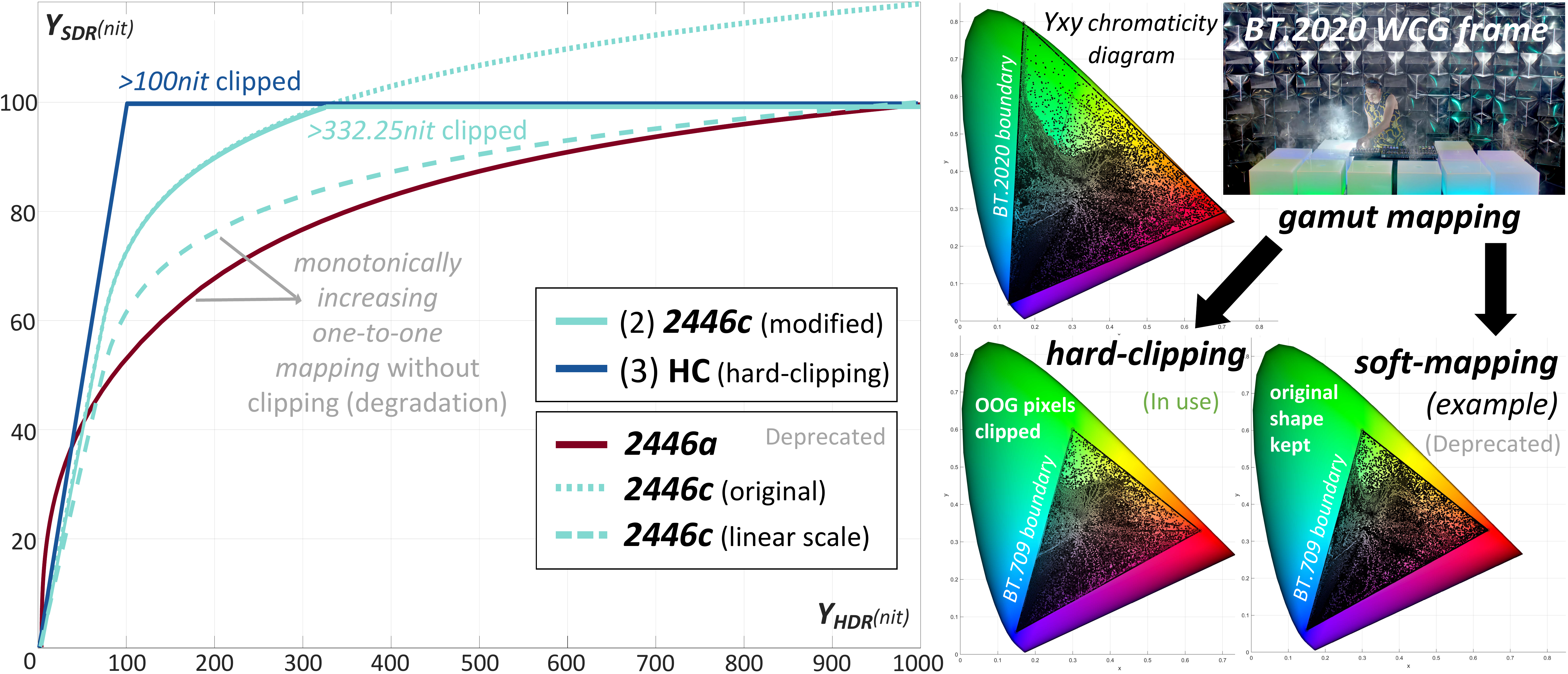}
	\caption{One of the key ideas of the proposed degradation models (DMs, \S \ref{sec:training_dm}): HDR's luminance (left) and color (right) volume should be properly clipped, so that network will learn corresponding restoration. Another idea is keeping a sensible brightness \& color/saturation \textit{style} during degradation, please refer to Tab.\ref{tab:sdr_stat}.}
	\label{fig:dm_curve}
\end{figure}

(2) \textit{\textbf{2446c+GM}}: Modified \textit{Method C} tone-mapping from Rec.2446\cite{BT2446}, followed by Gamut Mapping.
Here, each HDR encoded value $\mathbf{E}^{\prime}=[R^{\prime},G^{\prime},B^{\prime}]^{\mathrm{T}}$ (we use superscript $^{\prime}$ for non-linearity) is linearize to $\mathbf{E}$ by PQ EOTF, and then transferred to $Yxy$.
Then $Y$ is globally mapped to $Y_{SDR}$, while $xy$ are kept and then recomposed with $Y_{SDR}$ to form $\mathbf{E}_{SDR}$ and later $\mathbf{E}^{\prime}_{SDR_{709}}$. Our modification lies in:
\begin{equation}
	Y_{SDR} = clamp(TM_{2446c}(Y),0,100nit) 
	\label{eq:curve_2446c}
\end{equation}
where $TM_{2446c}(\cdot)\in[0,118nit]$ is the original global mapping curve in \cite{BT2446} (Fig.\ref{fig:dm_curve} dashed line).
That is, we clip the original output range to $100nit$ to produce more \textbf{FOEP}, rather than linear-scale which share same shortcoming with DM \textit{\textbf{2446a}}.
Gamut mapping will be introduced in Eq.\ref{eq:gm_cst}\&\ref{eq:gm_hc}.

(3) \textit{\textbf{HC+GM}}: Hard-Clipping and Gamut Mapping, a container conversion with SDR part unchanged and all luminance/color in HDR/WCG volume hard-clipped. For luniannce part, we clip all >$100nit$ pixels to $100nit$:
\begin{equation}
	\mathbf{E}_{SDR} =10 \times clamp(\mathbf{E},0,0.1),\ \mathbf{E}\in[0,1]
	\label{eq:curve_hc}
\end{equation}
where $0.1$ correspond to $100nit$ in normalized $1000nit$ linear HDR $\mathbf{E}$, and $\times10$ is to adapt the container discrepancy (nominal peak $1$ means $1000nit\rightarrow100nit$).

So far, $\mathbf{E}_{SDR}$ is still in BT.2020 gamut. Therefore, we append Gamut Mapping (\textbf{+GM}): First, Color Space Transform (CST) from BT.2020 RGB primaries to BT.709:
\begin{equation}
	\setlength{\arraycolsep}{0.5pt}
	\mathbf{E}_{SDR_{OOG}}=\mathbf{M}\mathbf{E}_{SDR},\  \mathbf{M}=\left[\begin{tiny}
		\begin{array}{ccc}
			1.6605 & 0.5876 & -0.0728\\
			-0.1246 & 1.1329 & 0.0083\\
			-0.0182 & -0.1006 & 1.1187
		\end{array}
	\end{tiny}\right]
	\label{eq:gm_cst}
\end{equation}
where $\exists\ \mathbf{E}_{SDR_{OOG}} \notin [0,1]$ \ie out-of-gamut (OOG) pixels will fall outside valid range after CST to a narrower volume. Dealing OOG is the key of gamut mapping, instead of soft-mapping\cite{BT2407,GMBook} (used by \cite{GamutNet}'s DM) preserving as much WCG information to narrow gamut, we use hard-clipping which clips all OOG pixels to BT.709 boundary (Fig.\ref{fig:dm_curve}):
\begin{equation}
	\mathbf{E}_{SDR_{709}}=clamp(\mathbf{E}_{SDR_{OOG}},0,1)
	\label{eq:gm_hc}
\end{equation}

Then, $\mathbf{E}_{SDR_{709}}$ from both (2) \& (3) is converted to SDR encoded value $\mathbf{E}^{\prime}_{SDR_{709}}$ by BT.1886\cite{BT1886} OETF (approximate \textit{gamma2.22}) and 8bit JPEG compression with QF=80.

Our philosophy lies in that most explicitly-defined operations can be streamlined to a $clamp(\cdot)$ function:
Only when luminance and color volume is clipped, multiple-to-one mapping occurs in DM, the trained network can learn corresponding recovery ability.
Also, \textbf{ALL} \& \textbf{ASL} in Tab.\ref{tab:sdr_stat} show that these DMs produce SDR with reasonable style, ensuring that network will not learn style deterioration.


\section{Experiments}
\label{sec:exp}

\textbf{Training detail}.
For each HDR frame randomly resized to [0.25x,1x], 6 patches sized $600\times600$ are cropped from random position. Then, each of 6$\times$3878 HDR patches is degraded to SDR by 1 of the 3 proposed DMs in \textbf{equal probability}, and again stored in JPEG with QF=75.

All LSN parameters are Kaiming initialized, optimized by $l_1$ loss and AdaM with learning rate starting from $2\times10^{-4}$ and decaying to half every $10^5$ iters till $4\times10^5$.

\subsection{Criteria and configuration}
\label{sec:exp_cri}

\textbf{Criteria.} Since the destination of SDR-to-HDRTV up-conversion is human perception, to provide a `step-change' improvement\cite{BT2381} of HDRTV than SDRTV, methods should be assessed based on if:
\textit{\textbf{C1:}} result is visually-pleasuring, from both brightness (\textit{\textbf{C1A}}) \& color appearance (\textit{\textbf{C1B}}) \etc,
\textit{\textbf{C2:}} HDRTV's advance on HDR/WCG volume is properly recovered and utilized,
\textit{\textbf{C3:}} bright and dark areas are recovered or at least enhanced and
\textit{\textbf{C4:}} artifacts are avoided.
Yet, as found by \cite{CheatHDR,SIHDRQA}, current distance-based metrics e.g. PSNR fails, since the main contributor of an appealing score lies in the accuracy of learned numerical distribution, rather \textit{\textbf{C1-4}}.
Hence, we use new assessment criteria in Tab.\ref{tab:assessment}.

\begin{table}[!h]
	\centering
	\scriptsize
	\begin{tabular}{|c|c|c|c|}
		\hline
		& \textbf{Visuals} in Fig.\ref{fig:teaser}\&\ref{fig:result}  & \textbf{Metrics} in Tab.\ref{tab:result} & \textbf{subj. exp.} (\S \ref{sec:exp_subj})                                                                                                      \\ \hline
		\textbf{\textit{C1A}} & \multirow{2}{*}{-} & \textbf{ALL}, \textbf{HDRBQ}        & \multirow{4}{*}{\begin{tabular}[c]{@{}c@{}}overall\\ rating \&\\ selection of\\ attribution\end{tabular}} \\ \cline{1-1} \cline{3-3}
		\textbf{\textit{C1B}} &                     & \textbf{ASL}               &                                                                                                            \\ \cline{1-3}
		\textbf{\textit{C3}}  & yellow\&blue boxes  & \multirow{2}{*}{-}                 &                                                                                                            \\ \cline{1-2}
		\textbf{\textit{C4}}  & detailed visuals    &                 &                                                                                                            \\ \hline
		\textbf{\textit{C2}}  & \textit{3D Yxy diagram}      & \textbf{FHL}(\textbf{WG})\textbf{P}, \textbf{EHL}(\textbf{WG}) & -                                                                                                          \\ \hline
	\end{tabular}
	\caption{How each criteria \textit{\textbf{C1-4}} is assessed: via \textbf{detailed visuals}, \textbf{fine-grained tailored metrics} and \textbf{subjective experiment}.}
	\label{tab:assessment}
\end{table}


\textbf{Competitors}. As in Tab.\ref{tab:result}, our method is compared with 6 learning-based
methods\cite{Kim19,Kim202,Zeng20,Chen211,Cao22,Xu222}, and 2 commercial software \textit{DaVinci} and \textit{Nuke}.
Results from joint-SR methods\cite{Kim19,Kim202,Zeng20} are \textit{bilinear} downscaled before comparison.
Note that learning-based methods are not re-trained with our dataset since we treat dataset as a crucial attribution to performance.

\textbf{Test set.} We select 12 4K SDR videos with 10$s$ duration and 50$fps$. 3 of 12 are degraded from HDR-GT counterpart by \textit{\textbf{YouTube}} DM (\S \ref{sec:hdrtv_dm}), the rest 9 are \textbf{\textit{graded}} from HDR sequence No. \textit{2},\textit{13},\textit{15},\textit{20},\textit{21},\textit{22},\textit{42},\textit{44},\textit{55} in \cite{BT2245}. Experiment fairness is guaranteed since both DMs (SDR-HDR relationship) in test set are `unseen' by our network (this even gives a bonus to methods trained with \textit{\textbf{YouTube}} DM\cite{Kim19,Kim202,Chen211,Cao22,Xu222}).

\subsection{Result}
\label{sec:exp_metrics}

\begin{figure*}[!h]
	\centering
	\includegraphics[width=0.9\linewidth]{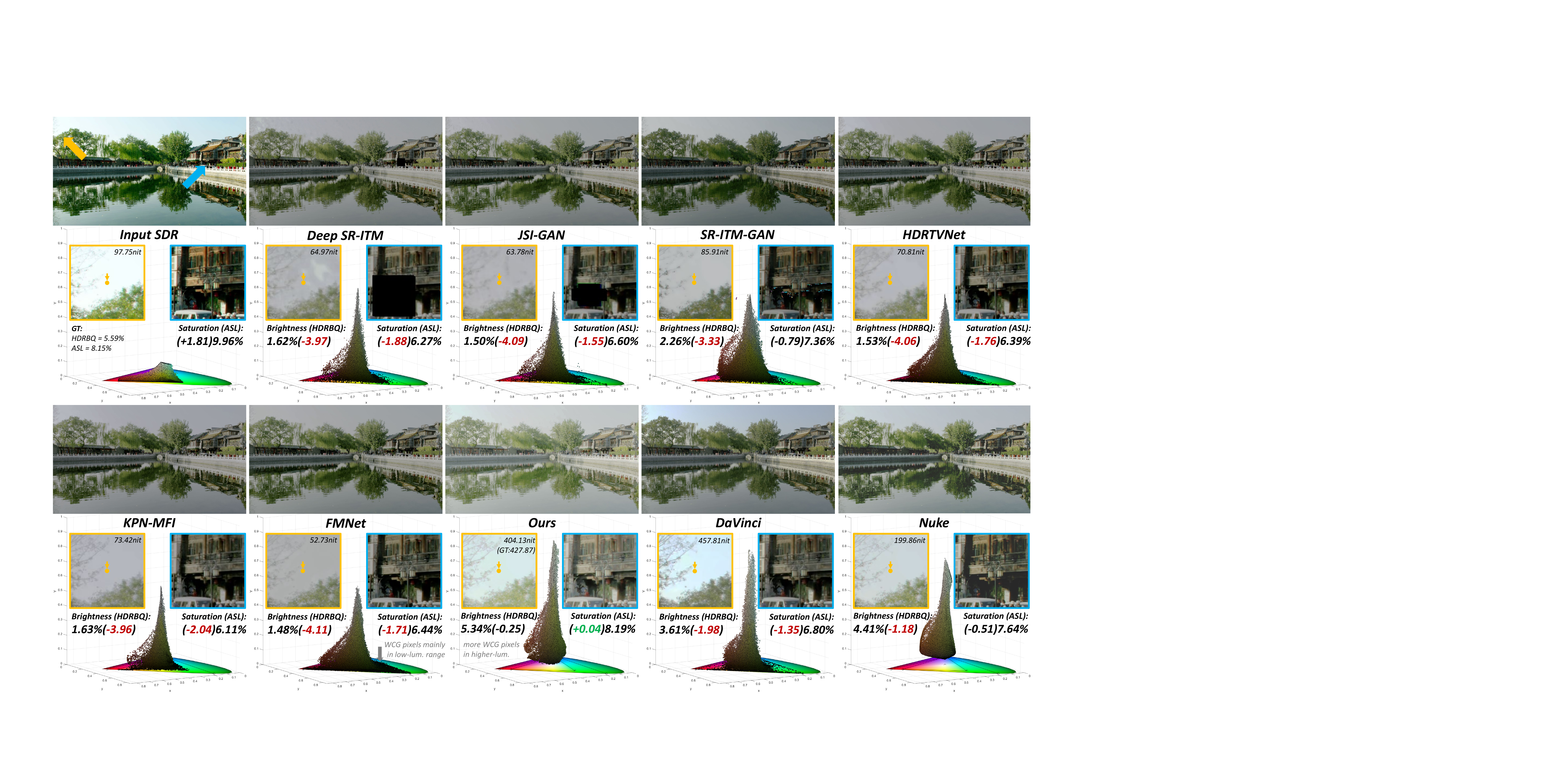}
	\caption{Visuals. We provide comparisons on \textit{3D Yxy chromaticity diagram} to assess how HDR/WCG volume is recovered (\textbf{\textit{C2}}), and detailed visuals on bright (yellow arrow, with luminance of same pixel indicated) and dark (blue arrow) areas to assess method's recover ability (\textbf{\textit{C3}}). Note that conclusion `SDR is more vivid than HDR' could not be drawn because HDR will appear dimmer than SDR in print version (explained in Fig.\ref{fig:teaser}). We hence turn to \textbf{ASL} \& \textbf{HDRBQ} (\textit{column 8-10}) and later subjective experiment to see if HDR is indeed dimmer (\textbf{\textit{C1}}). Comparison on more scenes will be provided in supplementary material.}
	\label{fig:result}
\end{figure*}

Results from all competitors are shown in Tab.\ref{tab:result} \& Fig.\ref{fig:result}.

\begin{table*}[!h]
	\centering
	\tiny
	\begin{tabular}{|lccrlrrrrrrrrrc|}
		\hline
		\multicolumn{1}{|c|}{\multirow{2}{*}{\begin{tabular}[c]{@{}c@{}}Method \\ (network) \end{tabular}}}    & \multicolumn{2}{c|}{\textit{how network is trained}}                                                              & \multicolumn{5}{c|}{(recovery rate \%, GT is 100\%) \textit{how HDR/WCG volume is recovered}}                                                             & \multicolumn{3}{c|}{(shift rate \%, GT is 0\%) \textit{overall-style}}                                     & \multicolumn{4}{c|}{\textit{conventional metrics}}                                                                    \\ \cline{2-15} 
		\multicolumn{1}{|c|}{}                            & \multicolumn{1}{c|}{dataset (GT)}                  & \multicolumn{1}{c|}{DM}                       & \multicolumn{2}{c}{\textbf{FHLP}}          & \multicolumn{1}{c}{\textbf{EHL}} & \multicolumn{1}{c}{\textbf{FWGP}}          & \multicolumn{1}{c|}{\textbf{EWG}}          & \multicolumn{1}{c}{\textbf{ASL}} & \multicolumn{1}{c}{\textbf{ALL}} & \multicolumn{1}{c|}{\textbf{HDRBQ}}         & \multicolumn{1}{c}{\textbf{PSNR}} & \multicolumn{1}{c}{\textbf{SSIM}} & \multicolumn{1}{c}{\textbf{$\Delta$E}} & \multicolumn{1}{c|}{\textbf{VDP3}} \\ \hline
		\multicolumn{1}{|l|}{Input SDR}                   & \multicolumn{2}{c|}{-}                                                                        & \multicolumn{2}{r}{0}             & 0                       & 0                                 & \multicolumn{1}{r|}{0}            & 6.570                   & 10.76                   & \multicolumn{1}{c|}{-}             & \multicolumn{1}{c}{23.92dB}    & \multicolumn{1}{c}{0.8861}    & \multicolumn{1}{c}{44.97}      & \multicolumn{1}{c|}{6.571}    \\ \cline{1-3}
		\multicolumn{1}{|l|}{Deep SR-ITM\cite{Kim19}}                 & \multicolumn{1}{c|}{\multirow{2}{*}{KAIST}}   & \multicolumn{1}{c|}{\multirow{2}{*}{\textbf{\textit{YouTube}}}} & \multicolumn{2}{r}{\cellcolor[HTML]{FFEEEE}{(14.03)0.2323}} & \cellcolor[HTML]{FFEEEE}{(9.70)0.372}             & (175.2)1.0964                     & \multicolumn{1}{r|}{(71.96)0.172} & \cellcolor[HTML]{FFEEEE}{(-20.33)5.485}           & \cellcolor[HTML]{FFEEEE}{(-15.06)9.580}           & \multicolumn{1}{r|}{\cellcolor[HTML]{FFEEEE}{(-72.80)1.428}} & 26.59dB                  & 0.8115                   & 32.54                      & 6.917                     \\ \cline{1-1}
		\multicolumn{1}{|l|}{JSI-GAN\cite{Kim202}}                     & \multicolumn{1}{c|}{}                         & \multicolumn{1}{c|}{}                         & \multicolumn{2}{r}{\cellcolor[HTML]{FFEEEE}{(12.33)0.2041}} & \cellcolor[HTML]{FFEEEE}{(3.55)0.136}             & (213.1)1.3334                     & \multicolumn{1}{r|}{(88.23)0.212} & \cellcolor[HTML]{FFEEEE}{(-16.62)5.741}           & \cellcolor[HTML]{FFEEEE}{(-14.36)9.659}           & \multicolumn{1}{r|}{\cellcolor[HTML]{FFEEEE}{(-79.20)1.092}} & 27.87dB                  & 0.8420                   & 30.23                      & 7.452                     \\ \cline{1-3}
		\multicolumn{1}{|l|}{SR-ITM-GAN\cite{Zeng20}}                  & \multicolumn{1}{c|}{Zeng20}                   & \multicolumn{1}{c|}{\textbf{\textit{Reinhard}}}                 & \multicolumn{2}{r}{\cellcolor[HTML]{FFEEEE}{(8.04)0.1332}}  & \cellcolor[HTML]{FFEEEE}{(14.33)0.550}            & \cellcolor[HTML]{FFEEEE}{(00.00)0.0000}                     & \multicolumn{1}{r|}{\cellcolor[HTML]{FFEEEE}{(00.00)0.000}} & \cellcolor[HTML]{FFEEEE}{(-7.04)6.400}            & \cellcolor[HTML]{FFEEEE}{(-30.37)7.853}           & \multicolumn{1}{r|}{\cellcolor[HTML]{FFEEEE}{(-57.12)2.251}} & 28.04dB                  & 0.8831                   & 24.78                      & 6.707                     \\ \cline{1-3}
		\multicolumn{1}{|l|}{HDRTVNet\cite{Chen211}}                  & \multicolumn{1}{c|}{\multirow{1}{*}{HDRTV1K}} & \multicolumn{1}{c|}{\multirow{3}{*}{\textbf{\textit{YouTube}}}} & \multicolumn{2}{r}{\cellcolor[HTML]{FFEEEE}{(18.59)0.3078}} & \cellcolor[HTML]{FFEEEE}{(16.29)0.625}            & (388.8)2.4334                     & \multicolumn{1}{r|}{\cellcolor[HTML]{FFEEEE}{(28.82)0.069}} & \cellcolor[HTML]{FFEEEE}{(-15.51)5.817}           & \cellcolor[HTML]{FFEEEE}{(-13.47)9.759}           & \multicolumn{1}{r|}{\cellcolor[HTML]{FFEEEE}{(-69.65)1.593}} & 30.82dB                  & 0.8812                   & 27.58                      & 8.120                     \\ \cline{1-2}
		\multicolumn{1}{|l|}{KPN-MFI\cite{Cao22}}                     & \multicolumn{1}{c|}{own}                      & \multicolumn{1}{c|}{}                         & \multicolumn{2}{r}{\cellcolor[HTML]{FFEEEE}{(1.17)0.0193}}  & \cellcolor[HTML]{FFEEEE}{(0.02)0.001}             & (392.9)2.4592                     & \multicolumn{1}{r|}{\cellcolor[HTML]{FFEEEE}{(15.82)0.038}} & \cellcolor[HTML]{FFEEEE}{(-21.74)5.388}          & \cellcolor[HTML]{FFEEEE}{(-16.10)9.462}           & \multicolumn{1}{r|}{\cellcolor[HTML]{FFEEEE}{(-82.47)0.920}} & 29.37dB                  & 0.8746                   & 27.47                      & 7.785                     \\ \cline{1-2}
		\multicolumn{1}{|l|}{FMNet\cite{Xu222}}                       & \multicolumn{1}{c|}{HDRTV1K}                  & \multicolumn{1}{c|}{}                         & \multicolumn{2}{r}{\cellcolor[HTML]{FFEEEE}{(13.67)0.2264}} & \cellcolor[HTML]{FFEEEE}{(12.35)0.474}            & (396.5)2.4813                     & \multicolumn{1}{r|}{(91.29)0.220} & \cellcolor[HTML]{FFEEEE}{(-16.91)5.770}           & \cellcolor[HTML]{FFEEEE}{(-13.48)9.758}           & \multicolumn{1}{r|}{\cellcolor[HTML]{FFEEEE}{(-71.24)1.510}} & 30.91dB                  & 0.8855                   & 27.16                      & 8.069                     \\ \cline{1-3}
		\multicolumn{1}{|l|}{\cellcolor[HTML]{EEFFEE}{LSN \textbf{(ours)}}}                  & \multicolumn{1}{c|}{\cellcolor[HTML]{EEFFEE}{HDRTV4K}}                  & \multicolumn{1}{c|}{\cellcolor[HTML]{EEFFEE}{\textbf{ours}$\times$3}}                  & \multicolumn{2}{r}{(256.6)4.2509} & (71.96)2.599            & (109.8)0.6873                     & \multicolumn{1}{r|}{(150.0)0.361} & (+9.25)7.522            & (+81.12)20.42           & \multicolumn{1}{r|}{(-27.04)3.829} & 24.47dB                  & 0.8310                   & 37.84                      & 8.130                     \\ \cline{1-3}
		\multicolumn{1}{|l|}{DaVinci}                     & \multicolumn{2}{c|}{\multirow{3}{*}{-}}                                                       & \multicolumn{2}{r}{(54.96)0.9103} & (43.16)1.655            & (310.0)1.9399                     & \multicolumn{1}{r|}{(85.22)0.205} & \cellcolor[HTML]{FFEEEE}{(-21.36)5.414}           & \cellcolor[HTML]{FFEEEE}{(-21.42)8.863}           & \multicolumn{1}{r|}{(-39.23)3.190} & 26.39dB                  & 0.8918                   & 35.47                      & 8.528                     \\ \cline{1-1}
		\multicolumn{1}{|l|}{Nuke}                        & \multicolumn{2}{c|}{}                                                                         & \multicolumn{2}{r}{(112.1)1.8565} & \cellcolor[HTML]{FFEEEE}{(24.70)0.384}            & \cellcolor[HTML]{FFEEEE}{(00.00)0.0000}                         & \multicolumn{1}{r|}{\cellcolor[HTML]{FFEEEE}{(00.00)0.000}}     & \cellcolor[HTML]{FFEEEE}{(-27.53)4.990}           & (+17.93)13.30           & \multicolumn{1}{r|}{\cellcolor[HTML]{FFEEEE}{(-51.19)2.562}} & 20.87dB                  & 0.7273                   & 64.29                      & 7.479                     \\ \cline{1-1} 
		\multicolumn{1}{|l|}{HDR-GT (\textcolor{blue}{ref.})}               & \multicolumn{2}{c|}{}                                                                         & \multicolumn{2}{r}{(\textcolor{blue}{100.0})1.6562} & (\textcolor{blue}{100.0})3.835            & (\textcolor{blue}{100.0})0.6258                     & \multicolumn{1}{r|}{(\textcolor{blue}{100.0})0.240} & (\textcolor{blue}{0.00})6.885             & (\textcolor{blue}{0.00})11.28             & \multicolumn{1}{r|}{(\textcolor{blue}{0.00})5.248}   & \multicolumn{1}{c}{-}    & \multicolumn{1}{c}{-}    & \multicolumn{1}{c}{-}      & \multicolumn{1}{c|}{-}    \\ \hline
		\multicolumn{15}{|c|}{\textbf{ablation studies $\downarrow$}}                                                                                                                                                                                                                                                                                                                                                                                                                                                    \\ \hline
		\multicolumn{1}{|l|}{\multirow{4}{*}{LSN \textbf{(ours)}}} & \multicolumn{1}{c|}{\multirow{1}{*}{\cellcolor[HTML]{EEFFEE}{HDRTV4K}}} & \multicolumn{1}{c|}{\cellcolor[HTML]{FFFFEE}{\textbf{\textit{YouTube}}}}                  & \multicolumn{2}{r}{\cellcolor[HTML]{FFEEEE}{(13.09)0.2168}}            & \cellcolor[HTML]{FFEEEE}{(6.63)0.254}                      & (401.3)2.5118                               & \multicolumn{1}{r|}{(59.81)0.144}           & \cellcolor[HTML]{FFEEEE}{(-14.68)5.874}                      & \cellcolor[HTML]{FFEEEE}{(-13.46)9.760}                      & \multicolumn{1}{r|}{\cellcolor[HTML]{FFEEEE}{(-76.89)1.213}}            & 30.15dB                       & 0.8858                       & 28.04                         & 7.902                        \\ \cline{2-3}
		\multicolumn{1}{|l|}{}                            & \multicolumn{1}{c|}{\multirow{1}{*}{\cellcolor[HTML]{EEFFEE}{HDRTV4K}}}       & \multicolumn{1}{c|}{\cellcolor[HTML]{FFFFEE}{\textbf{\textit{Reinhard}}}}                  & \multicolumn{2}{r}{\cellcolor[HTML]{FFEEEE}{(9.06)0.1501}}            & \cellcolor[HTML]{FFEEEE}{(19.14)0.734}                      & \cellcolor[HTML]{FFEEEE}{(00.00)0.0000}                                & \multicolumn{1}{r|}{\cellcolor[HTML]{FFEEEE}{(00.00)0.000}}           & (-5.74)6.490                      & \cellcolor[HTML]{FFEEEE}{(-25.65)8.387}                      & \multicolumn{1}{r|}{\cellcolor[HTML]{FFEEEE}{(-52.84)2.475}}            & 27.70dB                       & 0.8436                       & 26.03                         & 6.832                        \\
		\multicolumn{1}{|l|}{}                            & \multicolumn{1}{c|}{\cellcolor[HTML]{FFFFEE}{Zeng20}}                   & \multicolumn{1}{c|}{\cellcolor[HTML]{EEFFEE}{\textbf{ours}$\times$3}}                  & \multicolumn{2}{r}{(109.6)1.8147}            & \cellcolor[HTML]{FFEEEE}{(9.00)0.345}                      & \cellcolor[HTML]{FFEEEE}{(0.95)0.0593}                                & \multicolumn{1}{r|}{\cellcolor[HTML]{FFEEEE}{(0.76)0.002}}           & (+3.36)7.117                      & (+66.64)18.79                      & \multicolumn{1}{r|}{\cellcolor[HTML]{FFEEEE}{(-59.52)2.124}}            & 25.17dB                       & 0.8179                       & 28.58                         & 7.874                        \\
		\multicolumn{1}{|l|}{}                            & \multicolumn{1}{c|}{\cellcolor[HTML]{FFFFEE}{HDRTV1K}}                   & \multicolumn{1}{c|}{\cellcolor[HTML]{EEFFEE}{\textbf{ours}$\times$3}}                  & \multicolumn{2}{r}{(177.5)2.9405}            & (52.49)2.013                      & (279.5)1.7494                                & \multicolumn{1}{r|}{(70.43)0.169}           & (+6.56)7.337                      & (+52.30)17.18                      & \multicolumn{1}{r|}{(-34.83)3.420}            & 24.30dB                       & 0.8351                       & 38.03                         & 8.002                        \\ \hline
		
	\end{tabular}
	\caption{Metrics. We use \textbf{fine-grained tailored metrics} (\textit{column 3-9}, defined in Tab.\ref{tab:hdr-wcg-metrics}) to assess criteria \textbf{\textit{C1}}\&\textbf{\textit{2}} (\S \ref{sec:exp_cri}): The more significant metrics in \textit{column 3-6} are, the better HDR\&WCG volume is recovered (\textbf{\textit{C2}}) by specific method. Also, \textit{column 8-10} closer with GT stands for similar brightness\&color appearance. Note that we allow \textit{column 8-10} slightly bigger than GT, which means result HDR is reasonably more visual-pleasuring (\textbf{\textit{C1}}) than GT. Based on this, we highlight those results \colorbox[HTML]{FFEEEE}{unfavorable} for viewing experience.}
	\label{tab:result}
\end{table*}


\textbf{Metrics.}
From Tab.\ref{tab:result} we know that methods trained with \textbf{\textit{YouTube}} DM all tend to produce lower satiation (\textbf{ASL}) and brightness (\textbf{ALL}\&\textbf{HDRBQ}), and their \textbf{ASL} is lower even than input SDR.
Also, SR-ITM-GAN\cite{Zeng20} recovers least HDR\&WCG volume since its label HDR (from Zeng20 dataset) is of least \textit{extent of HDR/WCG} (Tab.\ref{tab:hdr_stat}).
Note that current methods produce adequate \textbf{FWGP}, but from \textit{3D Yxy chromaticity diagram} in Fig.\ref{fig:result} we known that their WCG pixels mainly exists in low luminance range, which means they are of little help to viewing experience.
Our method is able to recover adequate HDR/WCG volume, meanwhile reasonably enhance the brightness and saturation.
\footnote{We also provide conventional PSNR, SSIM, $\Delta$E (color difference \cite{BT2124}) and VDP3 (HDR-VDP-3\cite{VDP3}), but they mostly represent output's closer value with GT (For example, result both dimmer (\eg Deep SR-ITM\cite{Kim19}) and more vivid (ours) than GT will have a similar low score.), thus are helpless for our assessment. This phenomenon was first found by \cite{CheatHDR,SIHDRQA}, and will be further explained in supplementary material.} 

\textbf{Visuals.} Methods' recover ability is illustrated by Fig.\ref{fig:result}: Current ones underperform in both bright (yellow) and dark (blue) areas.
Specifically, methods trained with \textbf{\textit{YouTube}} DM produce dim (\textbf{HDRBQ}) and desaturated (\textbf{ASL}) result, and even get lower luminance than input SDR at same position (arrow in yellow box).
This confirms the finding in \S \ref{sec:hdrtv_dm} that network will learn to darken and desaturate if DM tend to brighten and over-saturate. Also, our method recovers more detail in bright and dark areas with a better style.

\subsection{Subjective experiment}
\label{sec:exp_subj} 

\begin{figure}[!h]
	\centering
	\includegraphics[width=0.75\linewidth]{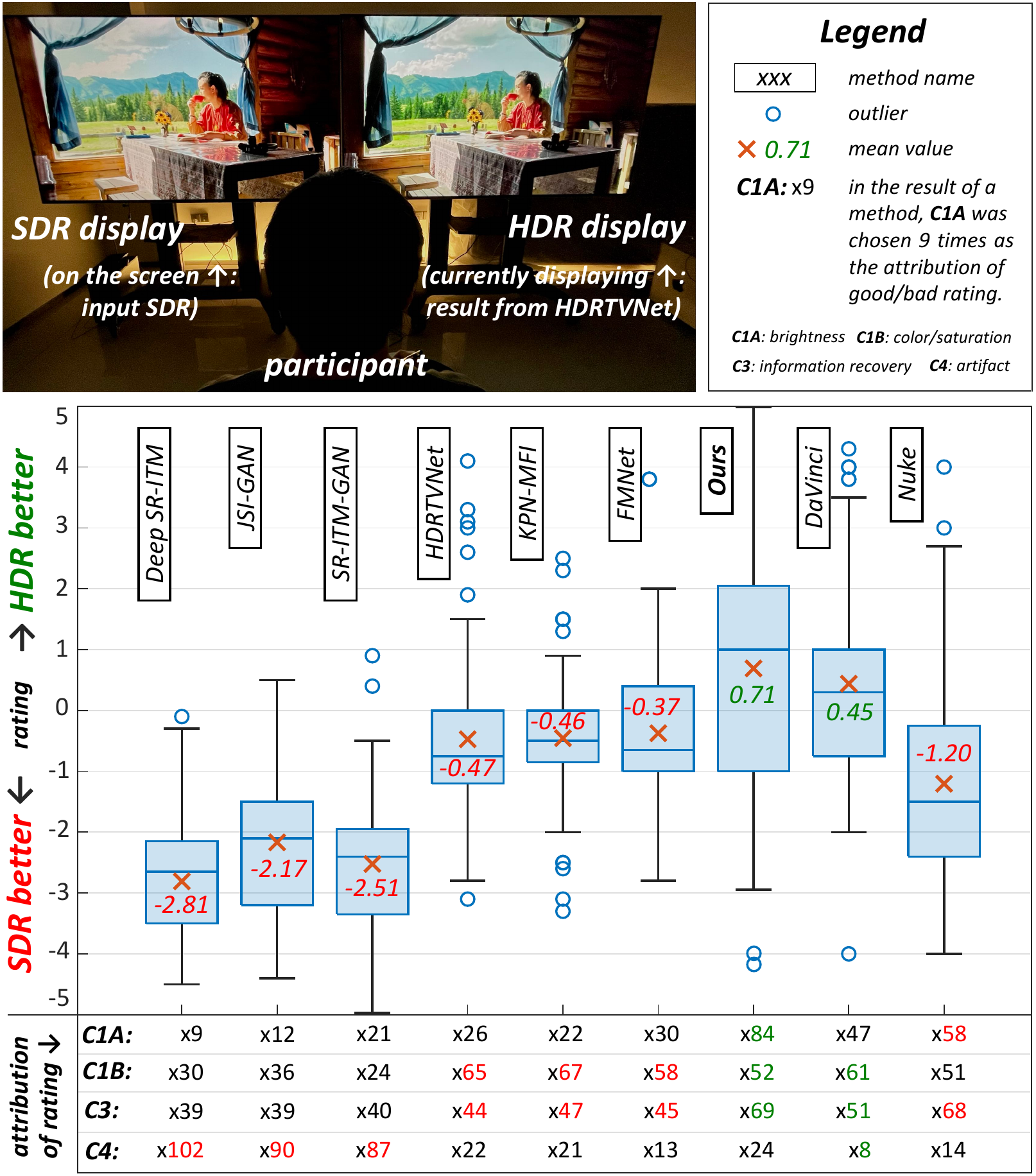}
	\caption{Environment and result of subjective experiment. 2 displays are caliberated differently to ensure both SDR and HDR are \textbf{correctly visualized}. Results are provided in quartile chart.}
	\label{fig:subj}
\end{figure}

Currently, few subjective studies\cite{Bist17,Luzardo20,Stojkovic21,Kim201,Mohammadi20,SIHDRQA} are designed for SDR-to-HDR procedure (rather between different HDR).
Similar to \cite{SIHDRQA}, we judge if output HDR is better than origin SDR.
Specifically, as in Fig.\ref{fig:subj}, we use 2 \textit{side-by-side} SONY KD85X9000H display supporting both SDR and HDRTV, one is calibrated to 100$nit$/BT.709 SDR and another PQ1000$nit$/BT.2020 HDR. Each (input)SDR-(output)HDR pair is displayed twice with 3$s$ gray level interval and following 10$s$ for rating:  each participant is asked to continuously rate from -5(SDR much better) to 0(same) to 5(HDR much better), meanwhile select at least 1 attribution (bottom Fig.\ref{fig:subj}) of his/her rating.
Such process is repeated 9(\#participant)$\times$9(\#competitor)$\times$12(\#clip) times.

Result shows that only 2 methods (ours \& \textit{DaVinci}) are recognized better than input SDR.
For methods\cite{Kim19,Kim202,Zeng20}, their main `attribution' (bottom Fig.\ref{fig:subj}) is artifact (\textbf{\textit{C4}}, see Fig.\ref{fig:result} blue box).
For artifact-free `\textbf{\textit{YouTube}}-DM' methods\cite{Chen211,Cao22,Xu222}, they are rated slightly worse mainly for lower saturation (\textbf{\textit{C1B}}) and incapability in information recovery, consisting with Tab.\ref{tab:result} and Fig.\ref{fig:result}.
Form `attribution' we also notice that our method got better score mainly for viewing experience (\textbf{\textit{C1}}) and information recovery (\textbf{\textit{C3}}).
Also, our `bad cases' lies in recovered large dark area with intrinsic noise \etc amplified and uneliminated.
This is why we got more checks on \textbf{\textit{C4}}($\times$24) than commercial methods \textit{DaVinci} and \textit{Nuke}. 

\subsection{Ablation studies}
\label{sec:exp_abl}

\begin{figure}[!h]
	\centering
	\includegraphics[width=0.95\linewidth]{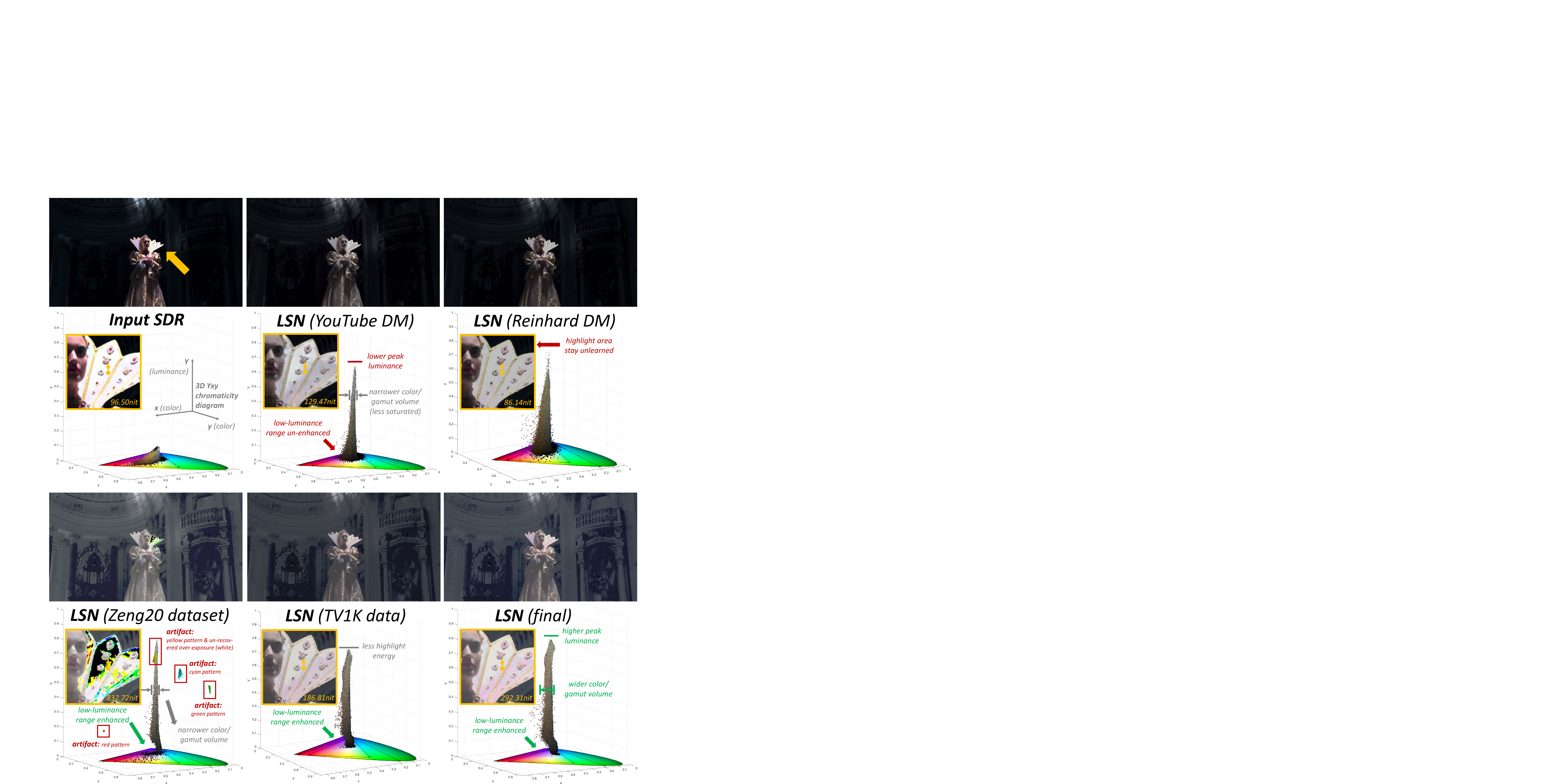}
	\caption{Result of ablation studies proves the importance of both high-quality label HDR and rational DMs. Specifically, absent of both label HDR's HDR/WCG volume (Zeng20) and DM's degradation (\textbf{\textit{Reinhard}}) will impair LSN's recover ability, meanwhile \textbf{\textit{YouTube}} DM's \textit{style} will make our LSN commonplace as others.}
	\label{fig:abl}
\end{figure}

\textbf{On DM.} When DM is changed to \textbf{\textit{YouTube}}, Tab.\ref{tab:result} witnesses a significant decline on \textbf{FHLP}, \textbf{EHL}, \textbf{ASL}, \textbf{ALL} and \textbf{HDRBQ}, while Fig.\ref{fig:abl} confirms a result similar to those `\textbf{\textit{YouTube}}-DM' methods, \ie worse viewing experience and less recover ability.
Also, when using \textbf{\textit{Reinhard}} DM which contains no clipping, result's highlight area stay unlearned.

\textbf{On dataset.} Here, we use original DMs, but label HDR from other dataset.
In Fig.\ref{fig:abl} yellow box, our DMs encourage the network to output higher luminance, but since Zeng20 is of least \textit{extent of HDR} \ie these highlight do not exist in label HDR, our LSN will not `recognize' them and thus produce artifact.
Since this dataset is also of least \textit{extent of WCG}, \textbf{FWGP}\&\textbf{EWG} in Tab.\ref{tab:result} drop obviously.
When using slightly-inferior HDRTV1K as label, difference is relatively less significant.
Yet, in both cases, \textbf{ASL}\&\textbf{ALL} are similar since DM \ie network's \textit{style} tendency is unaltered.

\section{Conclusion}
\label{sec:conclusion}

There are 2 types of low level vision: `sole-restoration' whose destination is only clean or GT \eg denoising, and `perceptual-motivated' aiming at better viewing experience \eg image enhancement/retouching.
SDR-to-HDRTV up-conversion belongs to both.
Yet, current methods only realize (it belongs to) the former and neglect the latter, leading their concentration only on network mechanism.

To this end, our response is two-fold:
\textbf{(1)} focusing on the impact of training set, and ameliorating its quality by proposing new dataset and DMs,
\textbf{(2)} involving novel assessment criteria based on the `perceptual' principal.

Results have proved the effectiveness of our ideas, and show that our method is ready for practical application.

{\small
\bibliographystyle{ieeetr}
\bibliography{egbib}

\begin{thebibliography}{10}

\bibitem{BT2381}
ITU, Geneva, Switzerland, {\em Report ITU-R BT.2381-0: Requirements for high
  dynamic range television (HDR-TV) systems}, 0~ed., 7 2015.

\bibitem{BT2100}
ITU, Geneva, Switzerland, {\em Recommendation ITU-R BT.2100-2: Image parameter
  values for high dynamic range television for use in production and
  international programme exchange}, 2~ed., 07 2018.

\bibitem{BT2020}
ITU, Geneva, Switzerland, {\em Recommendation ITU-R BT.2020-2: Parameter values
  for ultra-high definition television systems for production and international
  programme exchange}, 2~ed., 10 2015.

\bibitem{Kim19}
S.~Y. Kim, J.~Oh, and M.~Kim, ``Deep sr-itm: Joint learning of super-resolution
  and inverse tone-mapping for 4k uhd hdr applications,'' in {\em Proc. ICCV},
  pp.~3116--3125, 2019.

\bibitem{Kim202}
S.~Y. Kim, J.~Oh, and M.~Kim, ``Jsi-gan: Gan-based joint super-resolution and
  inverse tone-mapping with pixel-wise task-specific filters for uhd hdr
  video,'' in {\em Proc. AAAI}, vol.~34, pp.~11287--11295, 2020.

\bibitem{Zeng20}
H.~Zeng, X.~Zhang, Z.~Yu, and Y.~Wang, ``Sr-itm-gan: Learning 4k uhd hdr with a
  generative adversarial network,'' {\em IEEE Access}, vol.~8,
  pp.~182815--182827, 2020.

\bibitem{Chen211}
X.~Chen, Z.~Zhang, J.~S. Ren, L.~Tian, Y.~Qiao, and C.~Dong, ``A new journey
  from sdrtv to hdrtv,'' in {\em Proc. ICCV}, pp.~4500--4509, 2021.

\bibitem{Cao22}
G.~Cao, F.~Zhou, H.~Yan, A.~Wang, and L.~Fan, ``Kpn-mfi: A kernel prediction
  network with multi-frame interaction for video inverse tone mapping,'' in
  {\em Proc. IJCAI}, pp.~806--812, 2022.

\bibitem{Xu222}
G.~Xu, Q.~Hou, L.~Zhang, and M.-M. Cheng, ``Fmnet: Frequency-aware modulation
  network for sdr-to-hdr translation,'' in {\em Proc. ACMMM}, pp.~6425--6435,
  2022.

\bibitem{VDU}
Z.~Liang, ``Cvpr2022-1st workshop on vision dataset understanding.''
  \url{https://sites.google.com/view/vdu-cvpr22}, 2022.

\bibitem{Zhang21SRDM}
K.~Zhang, J.~Liang, L.~Van~Gool, and R.~Timofte, ``Designing a practical
  degradation model for deep blind image super-resolution,'' in {\em Proc.
  ICCV}, pp.~4791--4800, 2021.

\bibitem{Dewil22DNDM}
V.~Dewil, A.~Barral, G.~Facciolo, and P.~Arias, ``Self-supervision versus
  synthetic datasets: which is the lesser evil in the context of video
  denoising?,'' in {\em Proc. CVPR}, pp.~4900--4910, 2022.

\bibitem{Jiang21JPEGDM}
J.~Jiang, K.~Zhang, and R.~Timofte, ``Towards flexible blind jpeg artifacts
  removal,'' in {\em Proc. ICCV}, pp.~4997--5006, 2021.

\bibitem{Jiang22IEDM}
K.~Jiang, Z.~Wang, Z.~Wang, C.~Chen, P.~Yi, T.~Lu, and C.-W. Lin, ``Degrade is
  upgrade: Learning degradation for low-light image enhancement,'' in {\em
  Proc. AAAI}, vol.~36, pp.~1078--1086, 2022.

\bibitem{Zhou22UDCDM}
Y.~Zhou, Y.~Song, and X.~Du, ``Modular degradation simulation and restoration
  for under-display camera,'' in {\em Proc. ACCV}, pp.~265--282, 2022.

\bibitem{Guo2022SIHDR}
C.~Guo and X.~Jiang, ``Lhdr: Hdr reconstruction for legacy content using a
  lightweight dnn,'' in {\em Proc. ACCV}, pp.~3155--3171, 2022.

\bibitem{Restormer}
S.~W. Zamir, A.~Arora, S.~Khan, M.~Hayat, F.~S. Khan, and M.-H. Yang,
  ``Restormer: Efficient transformer for high-resolution image restoration,''
  in {\em Proc. CVPR}, pp.~5728--5739, 2022.

\bibitem{CheatHDR}
G.~Eilertsen, S.~Hajisharif, P.~Hanji, A.~Tsirikoglou, R.~K. Mantiuk, and
  J.~Unger, ``How to cheat with metrics in single-image hdr reconstruction,''
  in {\em Proc. ICCV}, pp.~3998--4007, 2021.

\bibitem{SIHDRQA}
P.~Hanji, R.~Mantiuk, G.~Eilertsen, S.~Hajisharif, and J.~Unger, ``Comparison
  of single image hdr reconstruction methods—the caveats of quality
  assessment,'' in {\em Proc. SIGGRAPH}, pp.~1--8, 2022.

\bibitem{HDRDNNSurvey}
L.~Wang and K.-J. Yoon, ``Deep learning for hdr imaging: State-of-the-art and
  future trends,'' {\em IEEE Trans. PAMI}, 2021.

\bibitem{IBL}
P.~Debevec, ``Image-based lighting,'' in {\em ACM SIGGRAPH 2006 Courses},
  pp.~4--es, 2006.

\bibitem{ReinhardHDRBook}
E.~Reinhard, W.~Heidrich, P.~Debevec, {\em et~al.}, {\em High dynamic range
  imaging: acquisition, display, and image-based lighting}.
\newblock Morgan Kaufmann, 2010.

\bibitem{Gharbi17HDRIE}
M.~Gharbi, J.~Chen, J.~T. Barron, S.~W. Hasinoff, and F.~Durand, ``Deep
  bilateral learning for real-time image enhancement,'' {\em ACM Trans.
  Graph.}, vol.~36, no.~4, pp.~1--12, 2017.

\bibitem{Yang18HDRIE}
X.~Yang, K.~Xu, Y.~Song, Q.~Zhang, X.~Wei, and R.~W. Lau, ``Image correction
  via deep reciprocating hdr transformation,'' in {\em Proc. CVPR},
  pp.~1798--1807, 2018.

\bibitem{Zheng21HDRIE}
Z.~Zheng, W.~Ren, X.~Cao, T.~Wang, and X.~Jia, ``Ultra-high-definition image
  hdr reconstruction via collaborative bilateral learning,'' in {\em Proc.
  ICCV}, pp.~4449--4458, 2021.

\bibitem{Mildenhall22HDRIE}
B.~Mildenhall, P.~Hedman, R.~Martin-Brualla, P.~P. Srinivasan, and J.~T.
  Barron, ``Nerf in the dark: High dynamic range view synthesis from noisy raw
  images,'' in {\em Proc. CVPR}, pp.~16190--16199, 2022.

\bibitem{Kalantari18MEHDR}
N.~K. Kalantari, R.~Ramamoorthi, {\em et~al.}, ``Deep high dynamic range
  imaging of dynamic scenes.,'' {\em ACM Trans. Graph.}, vol.~36, no.~4,
  pp.~144--1, 2017.

\bibitem{Wu18MEHDR}
S.~Wu, J.~Xu, Y.-W. Tai, and C.-K. Tang, ``Deep high dynamic range imaging with
  large foreground motions,'' in {\em Proc. ECCV}, pp.~117--132, 2018.

\bibitem{Yan19MEHDR}
Q.~Yan, D.~Gong, Q.~Shi, A.~v.~d. Hengel, C.~Shen, I.~Reid, and Y.~Zhang,
  ``Attention-guided network for ghost-free high dynamic range imaging,'' in
  {\em Proc. CVPR}, pp.~1751--1760, 2019.

\bibitem{Chen21MEHDR}
G.~Chen, C.~Chen, S.~Guo, Z.~Liang, K.-Y.~K. Wong, and L.~Zhang, ``Hdr video
  reconstruction: A coarse-to-fine network and a real-world benchmark
  dataset,'' in {\em Proc. ICCV}, pp.~2502--2511, 2021.

\bibitem{Niu21MEHDR}
Y.~Niu, J.~Wu, W.~Liu, W.~Guo, and R.~W. Lau, ``Hdr-gan: Hdr image
  reconstruction from multi-exposed ldr images with large motions,'' {\em IEEE
  Transactions on Image Processing}, vol.~30, pp.~3885--3896, 2021.

\bibitem{NTIRE22MEHDR}
E.~P{\'e}rez-Pellitero {\em et~al.}, ``Ntire 2022 challenge on high dynamic
  range imaging: Methods and results,'' in {\em Proc. CVPR}, pp.~1009--1023,
  2022.

\bibitem{Eilertsen17SIHDR}
G.~Eilertsen, J.~Kronander, {\em et~al.}, ``Hdr image reconstruction from a
  single exposure using deep cnns,'' {\em ACM Trans. Graph.}, vol.~36, no.~6,
  pp.~1--15, 2017.

\bibitem{Marnerides18SIHDR}
D.~Marnerides, T.~Bashford-Rogers, {\em et~al.}, ``Expandnet: A deep
  convolutional neural network for high dynamic range expansion from low
  dynamic range content,'' {\em Comput. Graph. Forum}, vol.~37, no.~2,
  pp.~37--49, 2018.

\bibitem{Liu20SIHDR}
Y.-L. Liu, W.-S. Lai, Y.-S. Chen, Y.-L. Kao, M.-H. Yang, Y.-Y. Chuang, and
  J.-B. Huang, ``Single-image hdr reconstruction by learning to reverse the
  camera pipeline,'' in {\em Proc. CVPR}, pp.~1651--1660, 2020.

\bibitem{Santos20SIHDR}
M.~S. Santos, T.~I. Ren, and N.~K. Kalantari, ``Single image hdr reconstruction
  using a cnn with masked features and perceptual loss,'' {\em ACM Trans.
  Graph.}, vol.~39, no.~4, pp.~80--1, 2020.

\bibitem{Chen21SIHDR}
X.~Chen, Y.~Liu, {\em et~al.}, ``Hdrunet: Single image hdr reconstruction with
  denoising and dequantization,'' in {\em Proc. CVPR}, pp.~354--363, 2021.

\bibitem{Ning18}
S.~Ning, H.~Xu, L.~Song, R.~Xie, and W.~Zhang, ``Learning an inverse tone
  mapping network with a generative adversarial regularizer,'' in {\em Proc.
  ICASSP}, pp.~1383--1387, 2018.

\bibitem{Hirao18}
K.~Hirao, Z.~Cheng, M.~Takeuchi, and J.~Katto, ``Convolutional neural network
  based inverse tone mapping for high dynamic range display using lucore,'' in
  {\em 2019 IEEE International Conference on Consumer Electronics (ICCE)},
  pp.~1--2, 2019.

\bibitem{Kim18}
S.~Y. Kim, D.-E. Kim, and M.~Kim, ``Itm-cnn: Learning the inverse tone mapping
  from low dynamic range video to high dynamic range displays using
  convolutional neural networks,'' in {\em Proc. ACCV}, pp.~395--409, 2018.

\bibitem{Xu191}
Y.~Xu, S.~Ning, R.~Xie, and L.~Song, ``Gan based multi-exposure inverse tone
  mapping,'' in {\em Proc. ICIP}, pp.~4365--4369, 2019.

\bibitem{Xu192}
Y.~Xu, L.~Song, R.~Xie, and W.~Zhang, ``Deep video inverse tone mapping,'' in
  {\em 2019 IEEE Fifth International Conference on Multimedia Big Data
  (BigMM)}, pp.~142--147, 2019.

\bibitem{Kim201}
D.-E. Kim and M.~Kim, ``Learning-based low-complexity reverse tone mapping with
  linear mapping,'' {\em IEEE Transactions on Circuits and Systems for Video
  Technology}, vol.~30, no.~2, pp.~400--414, 2019.

\bibitem{Zou20}
J.~Zou, K.~Mei, and S.~Sun, ``Multi-scale video inverse tone mapping with
  deformable alignment,'' in {\em 2020 IEEE International Conference on Visual
  Communications and Image Processing (VCIP)}, pp.~9--12, 2020.

\bibitem{Chen212}
T.~Chen and P.~Shi, ``An inverse tone mapping algorithm based on multi-scale
  dual-branch network,'' in {\em 2021 International Conference on
  Culture-oriented Science \& Technology (ICCST)}, pp.~187--191, 2021.

\bibitem{Xu221}
G.~Xu, Y.~Yang, J.~Xu, L.~Wang, X.-T. Zhen, and M.-M. Cheng, ``Joint
  super-resolution and inverse tone-mapping: A feature decomposition
  aggregation network and a new benchmark,'' 2022.

\bibitem{He221}
G.~He, K.~Xu, L.~Xu, C.~Wu, M.~Sun, X.~Wen, and Y.-W. Tai, ``Sdrtv-to-hdrtv via
  hierarchical dynamic context feature mapping,'' in {\em Proc. ACMMM},
  pp.~2890--2898, 2022.

\bibitem{He222}
G.~He, S.~Long, L.~Xu, C.~Wu, J.~Zhou, M.~Sun, X.~Wen, and Y.~Dai, ``Global
  priors guided modulation network for joint super-resolution and inverse
  tone-mapping,'' 2022.

\bibitem{Xu223}
K.~Xu, L.~Xu, G.~He, C.~Wu, Z.~Ma, M.~Sun, and Y.-W. Tai, ``Sdrtv-to-hdrtv
  conversion via spatial-temporal feature fusion,'' 2022.

\bibitem{Shao22}
T.~Shao, D.~Zhai, J.~Jiang, and X.~Liu, ``Hybrid conditional deep inverse tone
  mapping,'' in {\em Proc. ACMMM}, pp.~1016--1024, 2022.

\bibitem{Mustafa22}
A.~Mustafa, P.~Hanji, and R.~K. Mantiuk, ``Distilling style from image pairs
  for global forward and inverse tone mapping,'' in {\em The 19th ACM SIGGRAPH
  European Conference on Visual Media Production (CVMP)}, pp.~1--10, 2022.

\bibitem{Yao23}
M.~Yao, D.~He, X.~Li, Z.~Pan, and Z.~Xiong, ``Bidirectional translation between
  uhd-hdr and hd-sdr videos,'' {\em IEEE Transactions on Multimedia}, 2023.

\bibitem{Tang22}
R.~Tang, F.~Meng, and L.~Bai, ``Zoned mapping network from sdr video to hdr
  video,'' in {\em 2022 IEEE 5th Advanced Information Management, Communicates,
  Electronic and Automation Control Conference (IMCEC)}, vol.~5,
  pp.~1466--1472, 2022.

\bibitem{ST2084}
SMPTE, NY, USA, {\em ST 2084:2014 - SMPTE Standard - High Dynamic Range
  Electro-Optical Transfer Function of Mastering Reference Displays}, 2014.

\bibitem{BT709}
ITU, Geneva, Switzerland, {\em Recommendation ITU-R BT.709-6: Parameter values
  for the HDTV standards for production and international programme exchange},
  6~ed., 6 2015.

\bibitem{Cheng22ITMDM}
Z.~Cheng, T.~Wang, Y.~Li, F.~Song, C.~Chen, and Z.~Xiong, ``Towards real-world
  hdrtv reconstruction: A data synthesis-based approach,'' in {\em Proc. ECCV},
  pp.~199--216, 2022.

\bibitem{ReinhardTMO}
E.~Reinhard, M.~Stark, P.~Shirley, and J.~Ferwerda, ``Photographic tone
  reproduction for digital images,'' in {\em Proc. 29th SIGGRAPH},
  pp.~267--276, 2002.

\bibitem{BT2446}
ITU, Geneva, Switzerland, {\em Report ITU-R BT.2446-1: Methods for conversion
  of high dynamic range content to standard dynamic range content and
  vice-versa}, 1~ed., 3 2021.

\bibitem{Mohammadi20}
P.~Mohammadi, M.~T. Pourazad, and P.~Nasiopoulos, ``A perception-based inverse
  tone mapping operator for high dynamic range video applications,'' {\em IEEE
  Transactions on Circuits and Systems for Video Technology}, vol.~31, no.~5,
  pp.~1711--1723, 2020.

\bibitem{Marnerides21SIHDR}
D.~Marnerides, T.~Bashford-Rogers, and K.~Debattista, ``Deep hdr hallucination
  for inverse tone mapping,'' {\em Sensors}, vol.~21, no.~12, p.~4032, 2021.

\bibitem{Zhang21SIHDR}
Y.~Zhang and T.~Ayd{\i}n, ``Deep hdr estimation with generative detail
  reconstruction,'' in {\em Computer Graphics Forum}, vol.~40, pp.~179--190,
  2021.

\bibitem{EWG}
L.~Bai, Y.~Yang, and G.~Fu, ``Analysis of high dynamic range and wide color
  gamut of uhdtv,'' in {\em 2021 IEEE 5th Advanced Information Technology,
  Electronic and Automation Control Conference (IAEAC)}, vol.~5,
  pp.~1750--1755, 2021.

\bibitem{BT2407}
ITU, Geneva, Switzerland, {\em Report ITU-R BT.2407-0: Colour gamut conversion
  from Recommendation ITU-R BT.2020 to Recommendation ITU-R BT.709}, 0~ed., 10
  2017.

\bibitem{BT500}
ITU, Geneva, Switzerland, {\em Recommendation ITU-R BT.500-14: Methodologies
  for the subjective assessment of the quality of television images}, 14~ed.,
  10 2019.

\bibitem{CF}
D.~Hasler and S.~E. Suesstrunk, ``Measuring colorfulness in natural images,''
  in {\em Human vision and electronic imaging VIII}, vol.~5007, pp.~87--95,
  2003.

\bibitem{ICtCp}
Dolby, ``What is ictcp? - introduction.''
  \url{https://professional.dolby.com/siteassets/pdfs/ictcp_dolbywhitepaper_v071.pdf}.

\bibitem{HDRBQ}
S.~Ploumis, R.~Boitard, and P.~Nasiopoulos, ``Image brightness quantification
  for hdr,'' in {\em 2020 28th European Signal Processing Conference
  (EUSIPCO)}, pp.~640--644, 2021.

\bibitem{Arri-HDR}
Arri, ``Camera sample footage \& reference image.''
  \url{https://www.arri.com/en/learn-help/learn-help-camera-system/camera-sample-footage-reference-image},
  2022.

\bibitem{Netflix-HDR}
Netflix, ``Netflix open content.'' \url{https://opencontent.netflix.com/},
  2020.

\bibitem{HdM-HDR}
J.~Froehlich, S.~Grandinetti, B.~Eberhardt, S.~Walter, A.~Schilling, and
  H.~Brendel, ``Creating cinematic wide gamut hdr-video for the evaluation of
  tone mapping operators and hdr-displays,'' in {\em Digital photography X},
  vol.~9023, pp.~279--288, 2014.

\bibitem{YouTube-UGC-HDR}
Y.~Wang, S.~Inguva, and B.~Adsumilli, ``Youtube ugc dataset for video
  compression research,'' in {\em 2019 IEEE 21st International Workshop on
  Multimedia Signal Processing (MMSP)}, pp.~1--5, 2019.

\bibitem{LIVE-HDR-IQA}
Z.~Shang, J.~P. Ebenezer, A.~C. Bovik, Y.~Wu, H.~Wei, and S.~Sethuraman,
  ``Subjective assessment of high dynamic range videos under different ambient
  conditions,'' in {\em Proc. ICIP}, pp.~786--790, 2022.

\bibitem{t-SNE}
L.~Van~der Maaten and G.~Hinton, ``Visualizing data using t-sne.,'' {\em
  Journal of machine learning research}, vol.~9, no.~11, 2008.

\bibitem{OCIO2}
D.~Walker, C.~Payne, P.~Hodoul, and M.~Dolan, ``Color management with
  opencolorio v2,'' in {\em ACM SIGGRAPH 2021 Courses}, pp.~1--226, 2021.

\bibitem{GMBook}
J.~Morovi{\v{c}}, {\em Color gamut mapping}.
\newblock John Wiley \& Sons, 2008.

\bibitem{GamutNet}
H.~Le, T.~Jeong, A.~Abdelhamed, H.~J. Shin, and M.~S. Brown, ``Gamutnet:
  Restoring wide-gamut colors for camera-captured images,'' in {\em Color and
  Imaging Conference}, vol.~2021, pp.~7--12, 2021.

\bibitem{BT1886}
ITU, Geneva, Switzerland, {\em Recommendation ITU-R BT.1886-0: Reference
  electro-optical transfer function for flat panel displays used in HDTV studio
  production}, 0~ed., 3 2011.

\bibitem{BT2245}
ITU, Geneva, Switzerland, {\em Report ITU-R BT.2245-10: HDTV and UHDTV
  including HDR-TV test materials for assessment of picture quality}, 10~ed., 9
  2022.

\bibitem{BT2124}
ITU, Geneva, Switzerland, {\em Recommendation ITU-R BT.2124-0: Objective metric
  for the assessment of the potential visibility of colour differences in
  television}, 0~ed., 1 2019.

\bibitem{VDP3}
K.~Wolski, D.~Giunchi, {\em et~al.}, ``Dataset and metrics for predicting local
  visible differences,'' {\em ACM Trans. Graph.}, vol.~37, no.~5, pp.~1--14,
  2018.

\bibitem{Bist17}
C.~Bist, R.~Cozot, G.~Madec, and X.~Ducloux, ``Tone expansion using lighting
  style aesthetics,'' {\em Computers \& Graphics}, vol.~62, pp.~77--86, 2017.

\bibitem{Luzardo20}
G.~Luzardo, J.~Aelterman, H.~Luong, S.~Rousseaux, D.~Ochoa, and W.~Philips,
  ``Fully-automatic inverse tone mapping algorithm based on dynamic mid-level
  tone mapping,'' {\em APSIPA Transactions on Signal and Information
  Processing}, vol.~9, p.~e7, 2020.

\bibitem{Stojkovic21}
A.~Stojkovic, J.~Aelterman, H.~Luong, H.~Van~Parys, and W.~Philips,
  ``Highlights analysis system (hans) for low dynamic range to high dynamic
  range conversion of cinematic low dynamic range content,'' {\em IEEE Access},
  vol.~9, pp.~43938--43969, 2021.

\end{thebibliography}
}

\end{document}